\documentclass[english,aps,prl,twocolumn]{revtex4}
\usepackage[T1]{fontenc}
\usepackage[latin9]{inputenc}
\usepackage{amstext}
\usepackage{graphicx}
\usepackage{esint}

\makeatletter
\@ifundefined{textcolor}{}
{%
 \definecolor{BLACK}{gray}{0}
 \definecolor{WHITE}{gray}{1}
 \definecolor{RED}{rgb}{1,0,0}
 \definecolor{GREEN}{rgb}{0,1,0}
 \definecolor{BLUE}{rgb}{0,0,1}
 \definecolor{CYAN}{cmyk}{1,0,0,0}
 \definecolor{MAGENTA}{cmyk}{0,1,0,0}
 \definecolor{YELLOW}{cmyk}{0,0,1,0}
}

\usepackage[singlelinecheck=false,justification=raggedright]{caption}

\@ifundefined{showcaptionsetup}{}{%
 \PassOptionsToPackage{caption=false}{subfig}}
\usepackage{subfig}
\makeatother

\usepackage{babel}
\begin{document}

\title{The derivative discontinuity of the exchange-correlation functional}

\author{Paula Mori-Sánchez\footnote[1]{paula.mori@uam.es}}

\affiliation{Departamento de Química and Instituto de Física de la Materia Condensada
(IFIMAC), Universidad Autónoma de Madrid, 28049, Madrid, Spain}

\author{Aron J. Cohen\footnote[4]{ajc54@cam.ac.uk}}

\affiliation{Department of Chemistry, Lensfield Rd, University of Cambridge, Cambridge,
CB2 1EW, UK }
\begin{abstract}
The derivative discontinuity  is a key concept in electronic structure
theory in general and density functional theory in particular. The
electronic energy of a quantum system exhibits derivative discontinuities
with respect to different degrees of freedom that are a consequence
of the integer nature of electrons. The classical understanding refers
to the derivative discontinuity of the total energy as a function
of the total number of electrons ($N$), but it can also manifest
at constant $N$. Examples are shown in models including several Hydrogen
systems with varying numbers of electrons or nuclear charge ($Z$),
as well as the 1-dimensional Hubbard model (1DHM). Two sides of the
problem are investigated: first, the failure of currently used approximate
exchange-correlation functionals in DFT and, second, the importance
of the derivative discontinuity in the exact electronic structure
of molecules, as revealed by full configuration interaction (FCI).
Currently, all approximate functionals miss the derivative discontinuity,
leading to basic errors that can be seen in many ways: from the complete
failure to give the total energy of H$_{2}$ and H$_{2}^{+}$, to
the missing gap in Mott insulators such as stretched H$_{2}$ and
the thermodynamic limit of the 1DHM, or a qualitatively incorrect
density in the HZ molecule with two electrons and incorrect electron
transfer processes. Description of the exact particle behavior of
electrons is emphasized, which is key to many important physical processes
in real systems, especially those involving electron transfer, and
offers a challenge for the development of new exchange-correlation
functionals. 
\end{abstract}
\maketitle

\section{introduction}

The total energy in density functional theory (DFT) is given by
\begin{equation}
E[\rho]=T_{s}[\rho]+\int\!\!\rho(\mathbf{r})v_{\mathrm{ext}}(\mathbf{r})\mathrm{d}\mathbf{r}+J[\rho]+E_{xc}[\rho]\label{eq:KS energy}
\end{equation}
with explicit expressions for the non-interacting kinetic energy,
\begin{equation}
T_{s}[\rho]=\sum_{i}\langle\phi_{i}|-\frac{1}{2}\nabla^{2}|\phi_{i}\rangle,\label{eq:Ts}
\end{equation}
and Coulomb energy
\begin{equation}
J[\rho]=\frac{1}{2}\iint\!\!\frac{\rho(\mathbf{r})\rho(\mathbf{r}^{\prime})}{|\mathbf{r}-\mathbf{r}^{\prime}|}\mathrm{d}\mathbf{r}\mathrm{d}\mathbf{r}^{\prime}.\label{eq:Coulomb energy}
\end{equation}
All the unknown complexity and many-body physics are in the remaining
term, the exchange-correlation functional, $E_{xc}[\rho]$. The orbitals
used to construct the density are solutions of the Kohn-Sham equation
\begin{equation}
\left(-\frac{1}{2}\nabla^{2}+v_{\mathrm{ext}}(\mathbf{r})+\int\!\!\frac{\rho(\mathbf{r}^{\prime})}{|\mathbf{r}-\mathbf{r}^{\prime}|}\mathrm{d}\mathbf{r}^{\prime}+v_{xc}(\mathbf{r})\right)\phi_{i}(\mathbf{r})=\epsilon_{i}\phi_{i}(\mathbf{r})\label{eq:KS equation}
\end{equation}
where the exchange-correlation potential is given by the functional
derivative of the exchange-correlation energy, $v_{xc}(\mathbf{r})=\frac{\delta E_{xc}[\rho]}{\delta\rho(\mathbf{r})}$.
This is exact DFT. With the exact functional, the solution of the
Kohn-Sham equation (Eq. \ref{eq:KS equation}) to minimise the total
energy (Eq. \ref{eq:KS energy}) yields the exact solution of the
Schrödinger equation. However, the exact form of $E_{xc}[\rho]$ is
unknown and it is necessary to use density functional approximations
(DFA). DFAs have both an approximate energy expression, $E_{xc}^{{\rm DFA}}$,
and approximate Kohn-Sham potential, $v_{xc}^{\mathrm{DFA}}(\mathbf{r})$,
giving rise to approximate density, $\rho^{\mathrm{DFA}}(\mathbf{r})$,
and eigenvalues, $\{\epsilon_{i}^{\mathrm{DFA}}\}$.

\subsection{Density functional approximations}

There are many different functional forms, starting with semilocal
functionals that range from the local density approximation (LDA)
\cite{Dirac30,Vosko801200,Perdew9213244} to the generalized gradient
approximation (GGA) \cite{Becke883098,Lee88785,Perdew963865,Perdew868822,Handy01403}
and meta-GGA functionals \cite{Becke881053,Becke893761,Tao03146401,Zhao06194101}.
There are also many functionals that mix in Hartree-Fock exact exchange
in some manner, such as hybrid functionals with a varying degree of
constant admixture of exact exchange, from B3LYP (20\% HF) to PBE0
(25\%) to M06-2X (58\%) and M06-HF (100\%) \cite{Becke931372,Becke935648,Stephens9411623,Adamo996158,Becke978554,Hamprecht986264,Keal05121103}.
In the last decade many functionals have emerged that examine the
idea of range-separation pioneered by Andreas Savin, with functionals
that include all the long-range part of HF exchange (LC-BLYP, LC-$\omega$PBE)
to CAMB3LYP and $\omega$B97xd \cite{Iikura013540,Yanai0451,Vydrov06234109,Chai086615}
to mixing in only the short range part of Hartree-Fock \cite{Heyd038207,Heyd06219906}.
All these functionals use only the occupied orbitals and fit in a
general sense to the first four rungs of Jacob's Ladder of density
functional approximations \cite{Perdew011}. On the fifth rung there
have been some ideas  that use the unoccupied orbitals and eigenvalues
in functionals such as B2PLYP \cite{Grimme0634108}, which mix in
some MP2-like terms or the random phase approximation (RPA), for example
direct RPA (dRPA) \cite{RPA1,Furche08114105} which uses the Coulomb
only response in the adiabatic fluctuation dissipation theorem.

\begin{figure}
\includegraphics[scale=0.33]{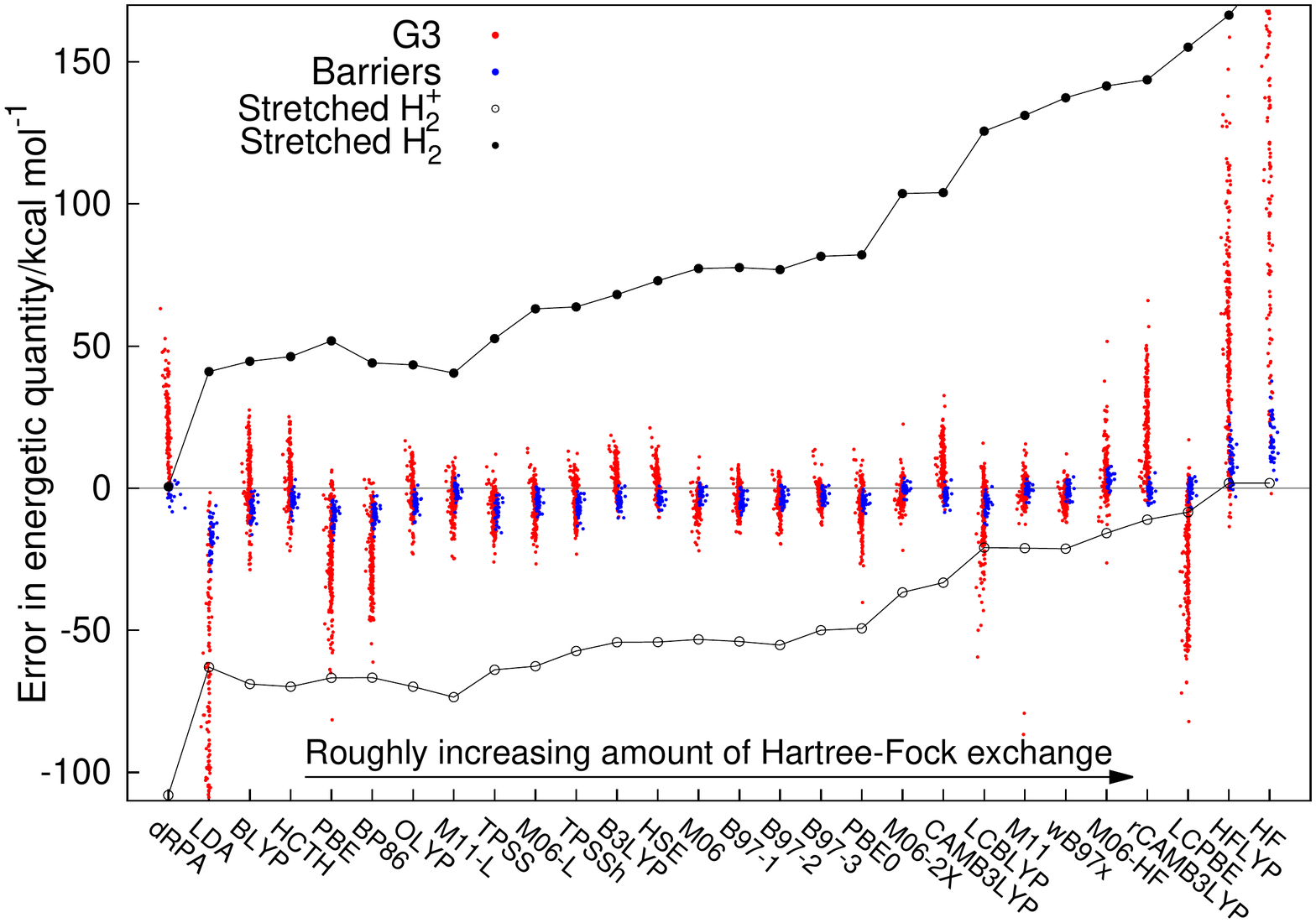}\caption{Calculated errors in thermochemical heats of formation (red dots),
reaction barrier heights (blue dots), and also the errors of stretched
H$_{2}^{+}$ (black circle) and stretched H$_{2}$ (black dots) for
a wide variety of functionals.\label{fig:G3-and-Barriers}}
\end{figure}

Fig. \ref{fig:G3-and-Barriers} presents a similar figure to Fig.
8 of Ref. \cite{Cohen12289} to illustrate the performance of a large
range of functionals for a set of thermochemical data (the heats of
formation of the G3 set \cite{Curtiss007374}) and a set of reaction
barrier heights \cite{Zhao042715,Zhao052012}. For each of the functionals
the calculated error of an energetic quantity for every individual
molecules is represented by a single dot, in Fig. \ref{fig:G3-and-Barriers}
each red dot is the error in the heat of formation of a molecule in
the G3 set and each blue dot is one of the errors in a reaction barrier
height. The results for dRPA for thermochemistry are only for the
G2 set \cite{Curtiss971063} and and the barriers are the DHB24 set
\cite{Zheng09808} and the results for these are taken from the paper
and supplementary information of Ref. \cite{Yang13174110} all other
calculations are post-B3LYP\cite{Cohen12289}. This allows one to
see the performance of many different functionals in a global manner.
In addition to the usual thermochemistry and barriers for each individual
functional we also plot the errors for the two simplest molecules
in the whole of chemistry infinitely stretched H$_{2}^{+}$ and infinitely
stretched H$_{2}$. Individually these molecules are known to be difficult
for functionals to describe with stretched H$_{2}^{+}$ \cite{Merkle929216}
epitomising self-interaction error \cite{Perdew815048,Mori-Sanchez08146401}
and stretched H$_{2}$ the problem of static correlation\cite{Cohen08121104}.
The errors for these two molecules, as one can see in Fig. \ref{fig:G3-and-Barriers},
are very large but importantly connected. One can see that in changing
functionals it is only possible to improve one error but with a corresponding
failure on the other, the two seem connected. No functional is able
to describe correctly these two simple molecules. It is this connection
between different systems that epitomises the challenge of making
one functional that can act discontinuously for different particle
numbers, which is a markedly different challenge to the usual atomization
energies of the G3 set or barrier heights. 

While there has been much improvement in the prediction of thermochemistry
and reaction barriers over many years, using many different ideas
in functionals, there is no functional that can reproduce the energy
of these two simple systems.This can be viewed in two ways: one, is
that the challenges of chemistry are not so related to the electronic
structure of stretched H$_{2}$/H$_{2}^{+},$ which has lead to the
concept that DFT works well as long as one does not stretch bonds. It
is hoped hope that these errors do not cause a problem in the systems
under study. However, the other view is that if current approximations are
not able to correctly describe these two simple systems, then it should
not be expected that for an unknown chemical they will give the correct
answer. The key is not to just focus on the system, but on the behavior
of the electrons themselves. The errors in stretched H$_{2}$/H$_{2}^{+}$
show a fundamental failure to correctly describe the electrons in
those molecules and, as such, the description of similar electronic
structure in many other systems will also fail. If these errors are
not corrected, the inconsistencies of functionals will continue to
dominate over the true behavior of electrons.

\subsection{Newer ideas in functionals}

There are several groups working on new ideas in DFT, which is greatly
needed to address the qualitative problems that can be seen in simple
model systems. For example, Gori-Giorgi and coworkers are looking
at the strictly-interacting $\lambda\rightarrow\infty$ limit of the
adiabatic connection for ideas based on strictly correlated electrons
(SCE) \cite{GoriGiorgi09166402,GoriGiorgi1014405,mirtschink_derivative_2013}.
The concept of SCE can deal with problem of stretched H$_{2}$. Other
notions beyond DFT, such as partition DFT, have been developed to
attempt to tackle some of these problems \cite{nafziger_delocalization_2013}.
Burke and coworkers have also looked at exact Kohn-Sham calculations
in 1D using the exact functional by doing a DMRG calculation via a
density perspective \cite{WSBW12}. The particle-particle RPA (ppRPA)
has been developed for electronic structure theory by van Aggelen,
Yang and Yang \cite{vanAggelen13030501}, showing a relation with
ladder coupled-cluster \cite{peng_equivalence_2013,ScuseriappRPACCSD}.
Becke also has used real space ideas to address non-dynamical correlation
and delocalization error using inverse hole models \cite{Becke0564101,becke_density_2013}.
There are schemes to combat this issue beyond DFT, from CASDFT \cite{Savin8859,Grafenstein00569}
to embedding methods in quantum chemistry \cite{FCIcode,FCIcode2,Huang11154110,libisch_origin_2012}
as well as ideas in density-matrix functional theory. Hopefully, the
culmination of these theories will provide new functionals to be able
to apply to a large spectrum of chemistry and physics without the
drawbacks of many of the currently used functionals.

\subsection{Challenge for DFAs}

In this paper we highlight the difficult question of the derivative
discontinuity of the exchange-correlation functional. The issue of
describing the energies of stretched H$_{2}^{+}$ and stretched H$_{2}$
with the same functional epitomises this challenge in a clear manner,
but there are many other ways to the view the problem. This is a much
larger issue than just that of stretching molecules, it is to correctly
describe the energy of the electrons in all situations. An improved
functional should be able to accurately describe the general behaviour
of electrons, especially in interesting physical processes where competing
effects act equally. This is the key problem to tackle, so that DFT
calculations can describe the important chemical reactions and responses
to electric and magnetic fields that are needed for the correct understanding
of the behaviour of electrons in enzyme catalysis, Li-ion batteries,
solar cells and many other technological applications.

\section{Derivative Discontinuity of the energy versus number of electrons}

The famous paper from Perdew, Parr, Levy and Balduz Jr in 1982 \cite{Perdew821691}
showed that the energy for a system with fractional electron number
is given by a straight line connecting integer electron numbers 
\begin{eqnarray}
E(N+\delta) & = & (1-\delta)E(N)+\delta E(N+1)\label{eq:PPLBEnergy}\\
\rho_{N+\delta}(\mathbf{r}) & = & (1-\delta)\rho_{N}(\mathbf{r})+\delta\rho_{N+1}(\mathbf{r}).
\end{eqnarray}
The energy and density are piecewise linear with straight lines connecting
the integer points. This means that, at the integers, both the energy
and density show (or can show) derivative discontinuities. In most
situations there is a large discontinuity in the density on changing
electron number. This is especially true in closed shell molecules
where the density difference between last electron added (given by
$\rho_{N}-\rho_{N-1})$ is very spatially different from where the
next electron is added $(\rho_{N+1}-\rho_{N}).$ Or, in other words,
when the frontier orbitals are spatially (and energetically) different. 

We will focus on understanding manifestations of the derivative discontinuity in
the total energy of integer systems. However, much of the understanding
of the DD is often related to the exact Kohn-Sham potential, which
was shown by Levy and Perdew \cite{Perdew831884} to undergo a jump
by a constant when passing through the integer. This constant, $C,$
is the derivative discontinuity.

\begin{eqnarray*}
v_{xc}^{N-\epsilon}(\mathbf{r}) & = & v_{xc}^{N}(\mathbf{r})\\
v_{xc}^{N+\epsilon}(\mathbf{r}) & = & v_{xc}^{N}(\mathbf{r})+C.
\end{eqnarray*}
This can be confusing to understand. For example, for a functional
such as LDA, what does it matter if the potential is shifted by a
constant? If that shifted potential is put in to the Kohn-Sham equations
Eq. \ref{eq:KS equation}, it will give rise to identical orbitals
and density, however the eigenvalues are shifted by a constant. If those
orbitals and density are put in to the energy expression Eq. \ref{eq:KS energy} an
identical energy will be obtained. This means that there is a discontinuous
change only in our eigenvalues, not in the total energy. This question
is part of the challenge of understanding the importance of exact
conditions in DFT, in this case the relation between the derivative
discontinuity and the energy expression in relation to the potential.
We think that the shift in the potential comes from something in the
total energy expression that is missing in LDA. It is this missing
part of the LDA energy expression that is the key question to be searching
for. This whole discussion is fraught with problems, but we want to
cement our understanding by finding molecules (or model systems) where
the question becomes clarified. In this work we will elaborate on the
implications of the derivative discontinuity for the energies of systems
with integer number of electrons, where there is no need to invoke
eigenvalues, fractional numbers of electrons or ensemble densities.

\subsection{Hydrogen atom and flat plane condition}

Consider the energy of a hydrogen atom, going from H$^{+}$ to H$^{-}$,
but with a possibly fractional number of electrons, $N=n_{\alpha}+n_{\beta}$,
$0\le n\le2$ and also $n_{\alpha}\le1$ and $n_{\beta}\le1$. This
is a very simple system to help understand fractional numbers of electrons.
The behavior of the exact energy is given by the the flat plane condition,
which is a very stringent test of approximate functionals \cite{Mori-Sanchez09}.
Especially important is the understanding of the behaviour at $N=1,$
where two planes intersect giving an energy derivative discontinuity,
and to consider adding and subtracting fractional numbers of electrons.
More complicated surfaces were also investigated in the work of Gal
and Geerlings \cite{Gal1032512}. 

DFAs really struggle to describe the flat plane and they completely
fail to recover the discontinuous behaviour seen in the exact behaviour
of the energy at $N=1$. It is this failure that is connected to (or the
root of) all the subsequent failures that we see. For the $n_{\alpha}=n_{\beta}$
line, approximate functionals massively fail, as they need to know
that on going from $E[0.4\alpha,0.4\beta]\rightarrow E[0.5\alpha,0.5\beta]\rightarrow E[0.6\alpha,0.6\beta]$
they have passed through one electron. Compare this with the edge
of the flat plane, $E[0.8\alpha,0\beta]\rightarrow E[\text{1.0\ensuremath{\alpha},0\ensuremath{\beta}]\ensuremath{\rightarrow}E[1.0\ensuremath{\alpha},0.2\ensuremath{\beta}]}$,
where there is a change of the orbital being occupied (from $\alpha$
to $\beta$) on passing through the integer. It is clear from the
energy expressions of Eq. \ref{eq:KS energy} that without a change
of orbitals, the only term that can give the discontinuity is the
exchange-correlation term. This is where the failure of current functionals
and the challenge for future functionals lie. There is also a large
discontinuity in the density with electron addition $f_{+}=\rho_{H^{-}}-\rho_{H}$
very spatially different from electron removal $f_{-}=\rho_{H}-\rho_{H^{+}}$.

\begin{figure}
\includegraphics[angle=-90,scale=0.35]{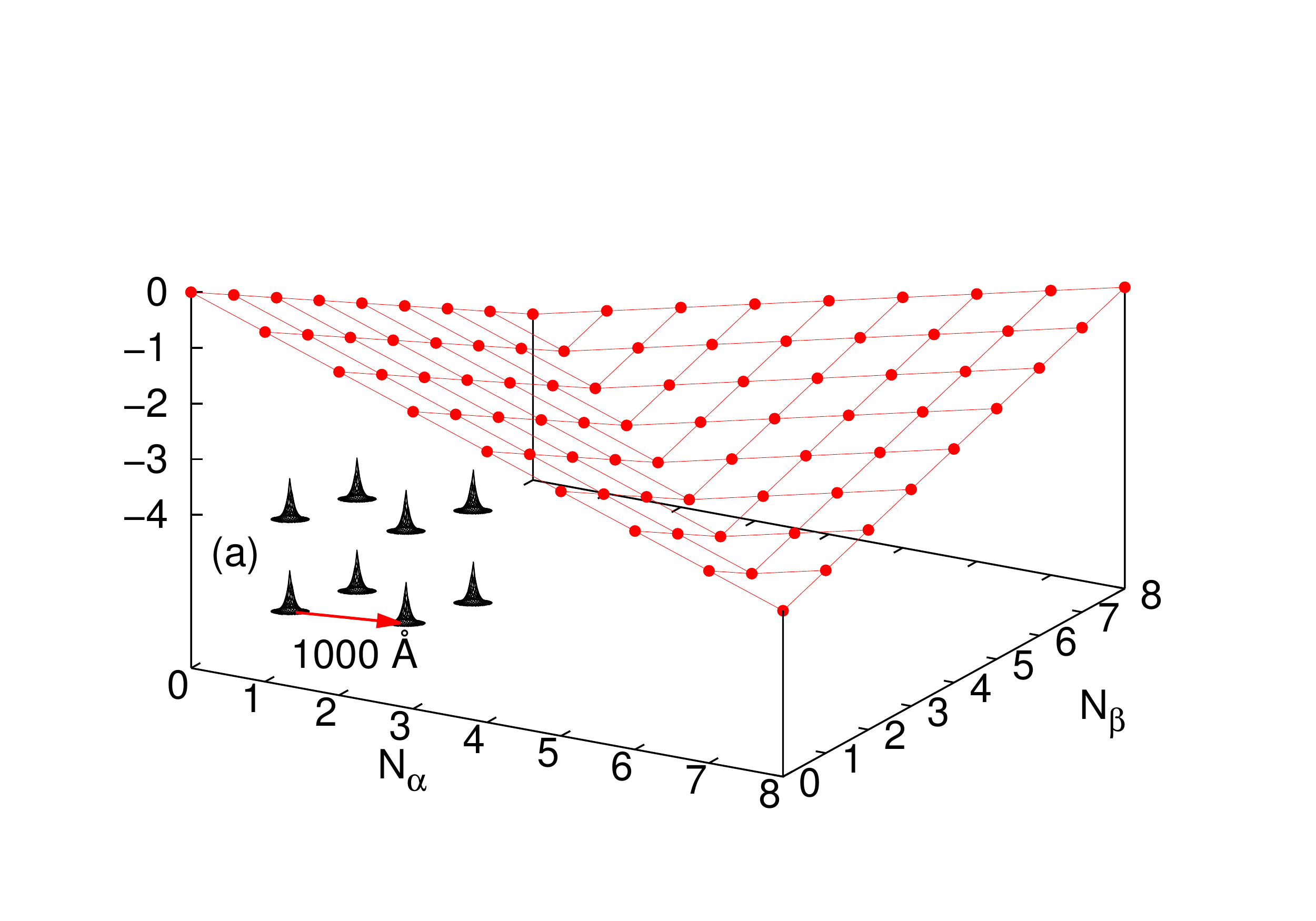}\caption{The total energy for different number of electrons of an H$_{8}$
cube separated by 1000 Å (sketched in inset (a)), with one basis function
per Hydrogen, calculated by FCI with up to $8\alpha$ and $8\beta$
electrons. \label{fig:H8 1basis}}
 
\end{figure}

\subsubsection{Hydrogen atom with 1 basis function}

To simplify the argument, it is useful to consider the calculation
of the Hydrogen atom just using a single basis function. The main
reason to do this is that now the density is constrained by the basis
function to be completely determined up to a factor such that it is
no longer discontinuous on passing through $N=1$, with $\rho_{H^{-}}=2\rho_{H}$.
A secondary point is that the discontinuity in the energy is in this
case greatly enhanced if the basis function is chosen as $\eta({\bf r})=\frac{1}{\sqrt{\pi}}e^{-r}$
(note the discontinuity could be reduced slightly if the basis function
is chosen to be give the correct density at $N=2$, i.e. $\eta({\bf r})\propto\sqrt{\rho_{{\rm H}^{-}}({\bf r})}$).
Additionally, the use of one basis function provides a direct connection
to the Hubbard model where there is also one basis function per site.
One of the features of the flat plane condition is that it considers
fractional numbers of electrons, which can lead to conceptual confusion
as well as technical challenges in extending methods to fractional
numbers of electrons\cite{yang_extension_2013,Steinmann13074107}.
However, the flat plane condition was developed to explain the root
cause in functionals of a general problem that affects real systems
 and hence can be equally seen in integer systems. 

Let us consider a cube of 8 Hydrogen atoms (each with one basis function).
For this system FCI calculations can be easily carried out with different
numbers of electrons and spins as in Fig. \ref{fig:H8 1basis}, where
we have used a very large (1000 Å) distance along the edge of the
cube just for simplicity. The density at each H atom is constrained
by the basis set so an increase in energy happens when more than one
electron per site is added; this is exactly the same as in the Hubbard
model, where the on-site repulsion $U$ causes an increase in energy
past half-filling. Of course if the basis set allowed, these electrons
would be unbound from the molecule (or as in a real H atom they would
be much more diffuse to slightly lower in energy). These issues could
be circumvented by changing the nucleus to be a He atom so that the
attraction to the nucleus makes it much more favourable in energy. The
real challenge for functionals it is to give the line of discontinuity
crossing 8 electrons (one electron per H atom).

\begin{figure}
\includegraphics[scale=0.32]{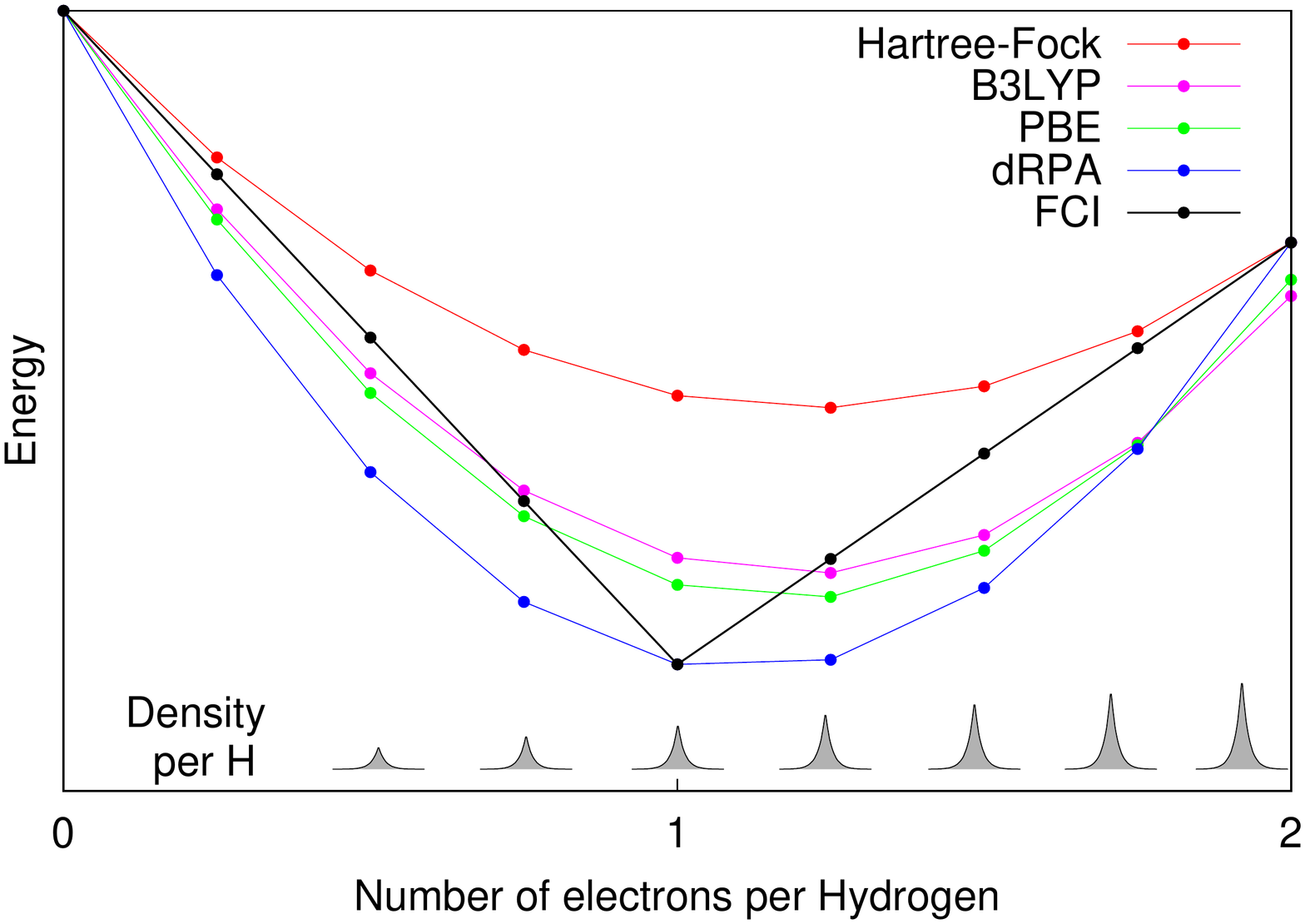}\caption{Energy of closed shell H$_{8}$ with a minimal basis set with 0, 2,
4, 6, 8, 10, 12, 14 and 16 electrons. Methods such as RHF, dRPA, BLYP
and B3LYP completely miss any discontinuous behaviour in the total
energy. \label{fig:H8 closed shell line }}
\end{figure}

\begin{figure*}[t]
\subfloat[Energy of the two site Hubbard model with different numbers of electrons
calculated with FCI, HF, MP2 and dRPA. $N=1$ is like a stretched
H$_{2}^{+}$ molecule and $N=2$ is like a stretched H$_{2}$ and
we can see how the errors of methods such as HF, dRPA and MP2 lead
to an incorrect gap at $N=2$.\label{fig:Hubbard2}]{\includegraphics[angle=-90,scale=0.33]{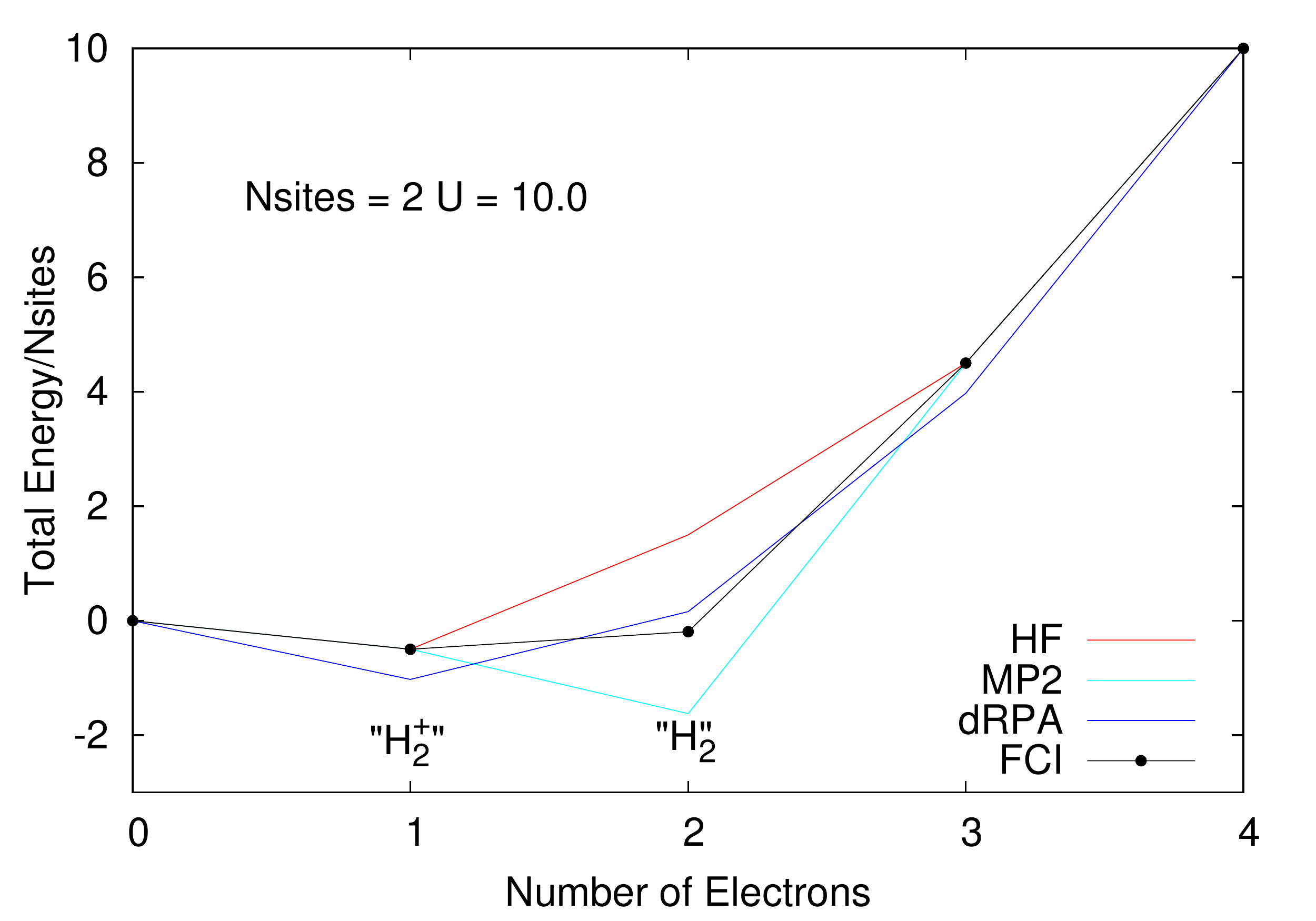}}\subfloat[Energy of a 50 site 1-d periodic Hubbard model with between 0 and
100 electrons calculated with HF, MP2 and dRPA compared with the exact
Bethe-Ansatz result. In contrast to the exact Bethe-Ansatz result
the DFA methods completely fail to give any sort of discontinuous
behaviour at half-filling (50 electrons).\label{fig:Hubbard50}]{\includegraphics[angle=-90,scale=0.33]{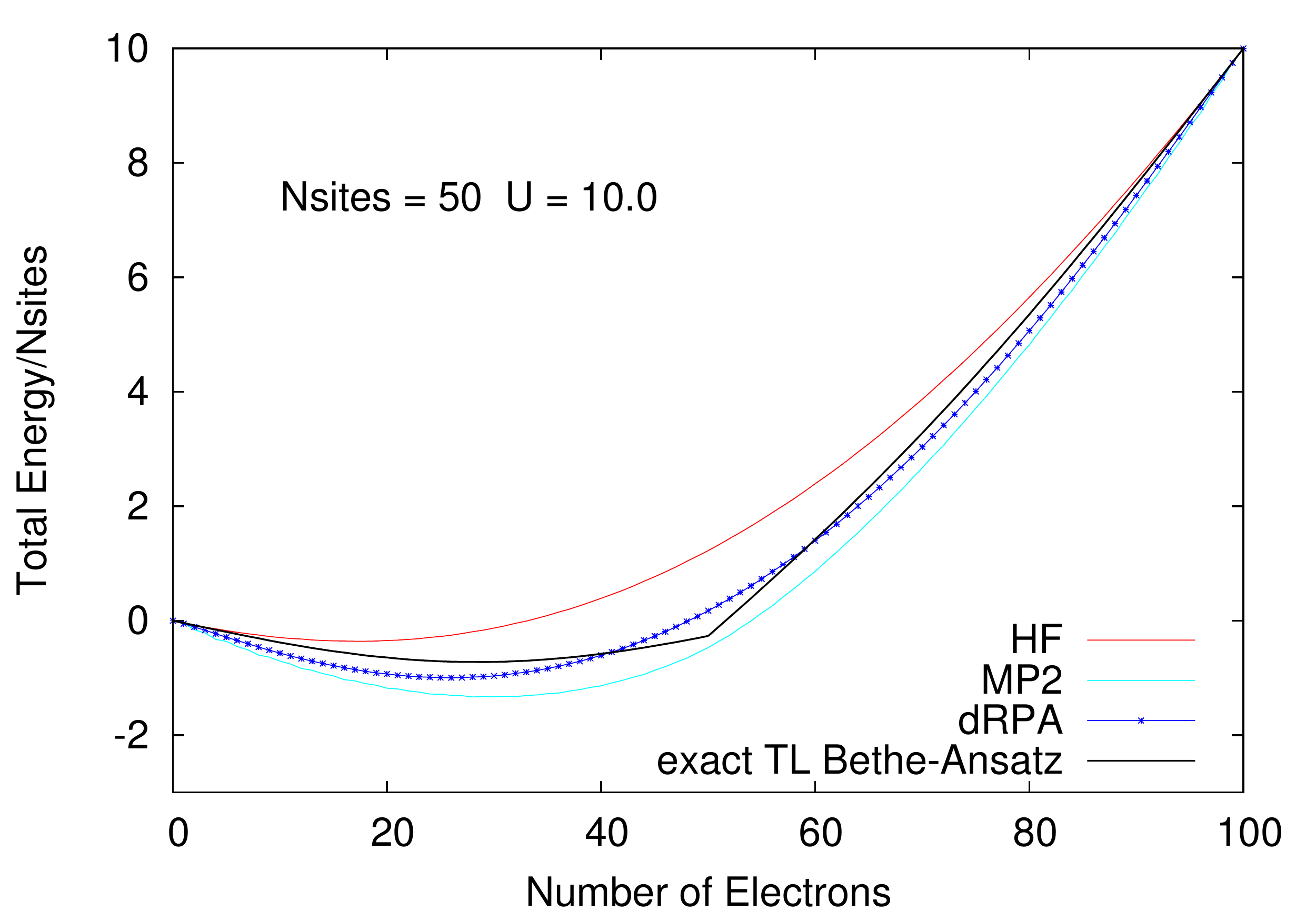}}\caption{\label{fig:Hubbard-Model}}
\end{figure*}

Let us just consider a single line in the flat plane with $N_{\alpha}=N_{\beta}$,
i.e. only closed shell systems. For this line calculations with any
functional can be easily carried out as the density and orbitals are
known. Fig. \ref{fig:H8 closed shell line } shows the performance
of HF, dRPA, PBE, B3LYP and FCI. First, the FCI energy with 16 electrons
(i.e. two electrons per site) is the same as HF, as the basis set
does not allow for any electron correlation. As DFT functionals such
as PBE and B3LYP treat correlation in a completely different manner,
they have a slightly lower energy at 16 electrons. We could have considered
just exchange functionals but have left in the correlation part so
that one can see the relatively small effect of dynamic correlation
functionals. 

Overall, all the approximate methods completely fail to reproduce
any discontinuous behaviour of the total energy, and have a smooth
behaviour in contrast to the correct answer of FCI. For less that
one electron per site the energy decreases by $-0.5$ per electron,
and for more than one electron per site the energy increases by $0.13$
per electron. This H$_{8}$ molecule clearly illustrates the same
physics as fractional electron numbers in one single H atom and also
the same error of functionals, the missing derivative discontinuity.
This example illustrates the important point that the missing derivative
discontinuity in functionals can manifest itself as an error in the
energy of integer systems.

\subsection{The 1-dimensional Hubbard model}

The Hubbard model \cite{Hubbard26111963} 
\[
\hat{H}=-t\sum_{i}(c_{i}^{\dagger}c_{i+1}+h.c.)+U\sum_{i}n_{i\alpha}n_{i\beta}
\]
is a much studied system in strongly correlated condensed matter physics
as it is a very simple model which describes interacting electrons
in narrow energy bands, and which has been applied to problems as
diverse as superconductivity, band magnetism, and the metal-insulator
transition. The interplay between the delocalized hopping, $t$, and
the localised repulsion, $U$, can lead to interesting balance in
physical behaviors. Here we will examine the 1-dimensional Hubbard
model (1D-HM) to highlight the connection with Hydrogen atoms and
derivative discontinuities, especially that seen at half filling.
For the 1D-HM the exact answer is known in the thermodynamic limit
using the Bethe-Ansatz, for example, the gap of the 1D-HM is given
by \cite{Lieb681445} 
\[
E_{{\rm gap}}^{{\rm BA}}=U-4+8\int_{0}^{\infty}\frac{J_{1}(x)}{x[1+\exp(Ux/2)]}\mathrm{d}x
\]
where $J_{1}$ is the first order Bessel function. 

The Hubbard model is  symmetric around half filling (except that above
half-filling an electron-interaction term is included) i.e. for a
system with 2$N$ sites and doping fraction of $M/N$
\begin{equation}
E[N-M]=E[N+M]+MU.\label{eq:Hubbard E symmetry}
\end{equation}
$ $This leads to a clear picture of the derivative discontinuity
that exists in the Hubbard model at half-filling, and raises the question
of how to include this physics in a functional. This is exactly the
same as the question of how to get a gap at the whole line in the
flat plane condition of the Hydrogen atom. The key of how to generally
put this in a functional is distinct from how to predict the gap of
the Hubbard model, where of course the knowledge of the system and
the property in Eq. (\ref{eq:Hubbard E symmetry}) can be more specifically
used to get the gap. For example, Capelle and coworkers \cite{CapelleHubbard,Franca12073021,CapelleReview}
have several specific functionals (BA-LDA (LSOC, FVC)) and different
parameterizations that are able to give the gap of the Hubbard model.

We consider finite Hubbard rings (chains with periodic boundary conditions)
where exact results can be computed using FCI. The size of the chain
can be increased and the large number of site limit corresponds to
the thermodynamic limit. The two site Hubbard model has a very clear
connection to infinitely stretched Hydrogen molecule, with one electron
this corresponds to H$_{2}^{+}$ and with two electrons to H$_{2}$. Fig.
\ref{fig:Hubbard-Model} shows the total energy for two different
Hubbard models, both with $t=1$ and a large values of $U=10$. The
two site model is examined in Fig \ref{fig:Hubbard2}. The performance
of methods is exactly as expected from calculations on stretched H$_{2}^{+}$/H$_{2};$
HF and MP2 are good for one electron but fail for two electrons (HF
is too high due to static correlation error and MP2 diverges to $-\infty$
as $U$ is increased). dRPA performs better for two electrons at the
cost of a massive error for one electron \cite{Mori-Sanchez12042507}.
Here, dRPA would still give a gap at half-filling ($N=2$). However,
as shown for the 50 site model in Fig. \ref{fig:Hubbard50}, the real
problem arises in approaching the thermodynamic limit, where methods
such as HF, dRPA or even MP2, smooth out and all semblance of a gap
at half-filling disappears. However, the exact Bethe-Ansatz results
give a very large gap. The complete failure of functionals to give
a derivative discontinuity means that their results on the Hubbard
model are completely physically incorrect as they miss one of the
key behaviours of true Hubbard electrons.

\subsection{Fractional nuclei: HZ$^{\{2e\}}$}

\begin{figure}[t]
\includegraphics[scale=0.31]{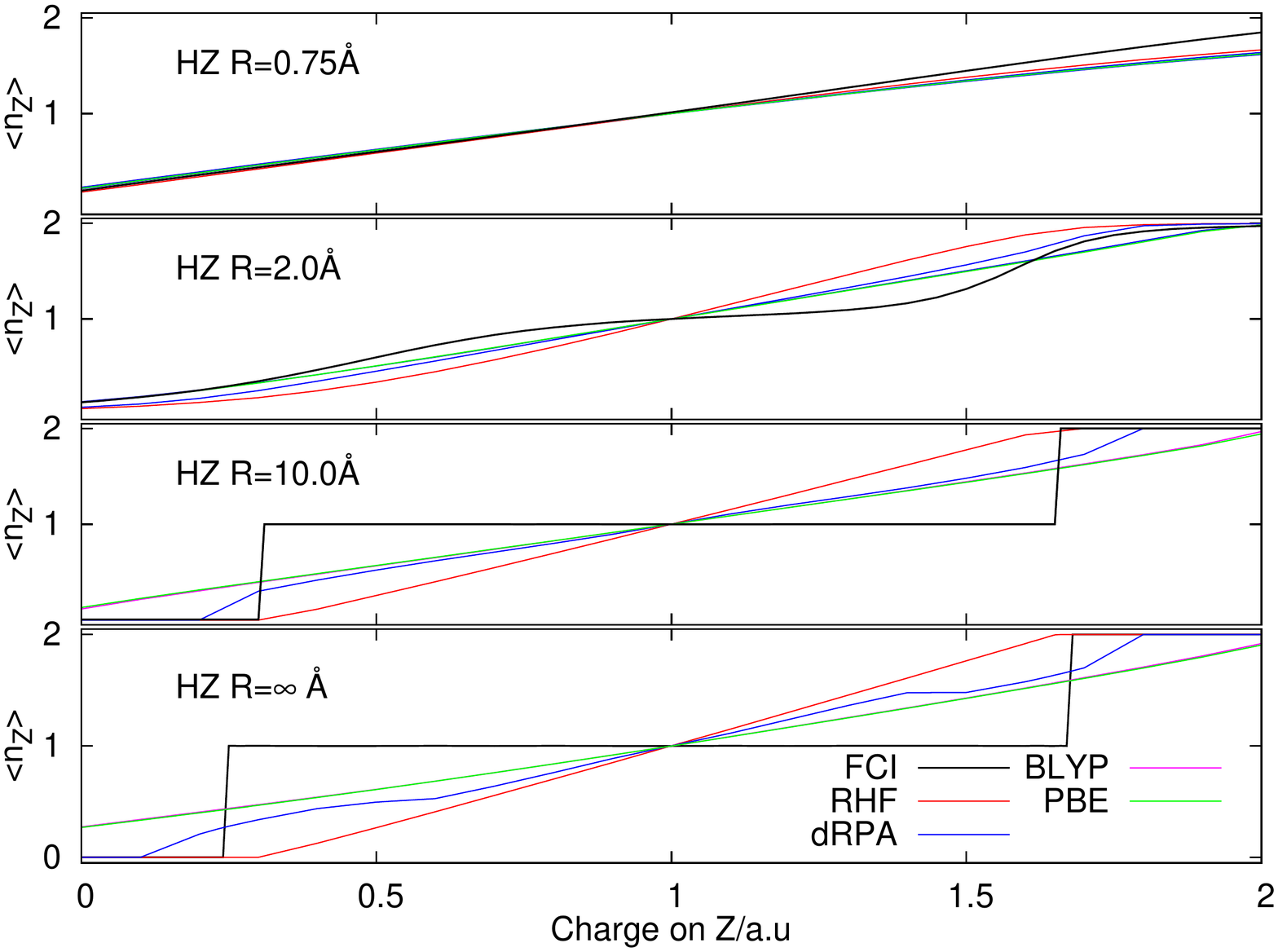}\caption{Occupation of Z atom, $\langle n_{Z}\rangle$, as the nuclear charge
on the Z atom is varied at four different bond lengths comparing FCI
and several different approximate DFT functionals.\label{fig:Occupation-HZ2e}}
\end{figure}

A picture of the discontinuous behaviour of the electrons is beautifully
illustrated in the two electron example of an H$_{2}$ like molecule
changing the nuclear charge of one of the protons  to be fractional,
giving HZ$^{\{2e\}}$ \cite{cohen_dramatic_2014}. This encompasses
a set of systems connected by a very simple change to the one-electron
potential. This is a smooth and continuous change to the Hamiltonian,
however, how does the electronic structure behave on these small changes,
does it also change smoothly? As demonstrated in Fig. \ref{fig:Occupation-HZ2e}
the answer is, of course, that it depends. In some cases, as illustrated
in short bond distances of HZ$^{\{2e\}}$, the electron also moves
smoothly on this change. However, at long distances, the electron
moves  discontinuously, being either on the H or the Z, it is not
shared between them.

The true behaviour of electrons, as given by FCI calculations, is
simple to understand for HZ$^{\{2e\}}$ at stretched bond lengths.
When $Z=0$ there are two electrons on the H atom (i.e. H$^{-})$
and as $Z$ increases an electron  moves from the H to the Z when
the energy of putting one electron on the Z atom gives a lower energy.
This occurs when the energy of one electron on the $Z$ atom, $E=-\frac{Z^{2}}{2}$,
is lower than the negative of the electron affinity of the H atom
(EA(H)=$-0.0277a.u.$). This happens at $Z=0.235$. The atoms are
too far apart to be bonded and no fractional electron transfer happens,
as can be understood from the PPLB straight line interpolation of
the true FCI energy of Eq. (\ref{eq:PPLBEnergy}). Something similar
happens around $Z=1.67$, where it becomes more energetically favourable
to have two electrons on the Z atom and none on the H (at that point
the electron affinity of the Z atom crosses $0.5$). This is simple
and clear, it is just the qualitative failure of functionals that
is surprising. 

Consider $Z=1,$ i.e. the H$_{2}$ molecule, all functionals get a
qualitatively correct density due to the symmetry of the problem.
However, they respond completely incorrectly to a small change in
one of the atoms. DFAs smoothly move electrons when in fact they should
not do so; no fractional charge is seen from FCI calculations. This
turns the well known static-correlation error in the energy into a
qualitative failure in the density. Approximate energy functionals
are not able to describe the integer nature of electrons and they
do not penalise correctly the splitting up of an electron in these
stretched cases. 

The HZ$^{\{2e\}}$ system is very interesting and has advantages over
very similar physics that can be seen in asymmetric Hubbard models
or Anderson models in that functionals can be simply tested and their
performance directly analyzed in real space. The conclusion is that
they all fail for this problem. The reasons can be traced back to
the failure of functionals for the closed-shell line of the flat plane
(see Fig. \ref{fig:H8 closed shell line }). Along that line these
functionals have no clear knowledge that they have passed through
one electron and this translates into the problems seen in the density
for these systems. This example is very illustrative but it is a slightly
hard test as it requires self-consistent determination of the density.
For methods such as dRPA, which is usually evaluated with PBE (sometimes
HF) orbitals, it requires a self-consistent calculation \cite{GGKS}
to highlight a problem in the density rather than energy, as carried
out for Fig. \ref{fig:Occupation-HZ2e} . 

\begin{figure}
\includegraphics[angle=-90,scale=0.33]{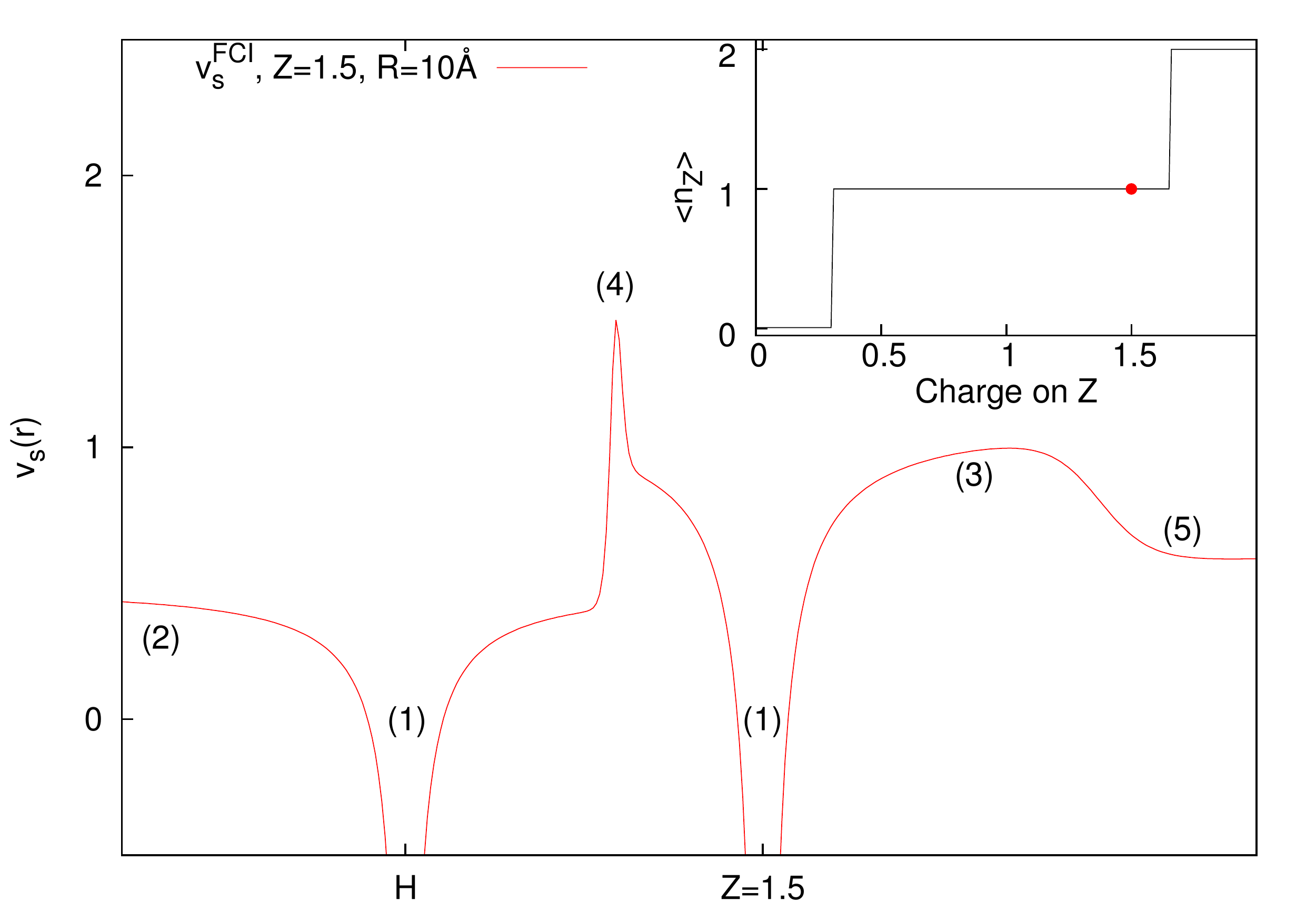}\caption{Kohn-Sham potential of HZ$^{\{2e\}}$ for $Z=1.5$ with a bond length
of 10 Å. The features of the potential given by labels (1)-(5) are
explained in the text and the inset  shows the number of electrons
on the Z atom for all HZ$^{\{2e\}}$ systems.\label{fig:Kohn-Sham-potential-HZ2e}}
\end{figure}

\section{The derivative discontinuity as a challenge for $v_{xc}(\mathbf{r})$?}

Our main view is that the derivative discontinuity must be understood
as a challenge for the energy, $E_{xc}[\rho]$. For example, in the
HZ$^{\{2e\}}$ problem, a functional that correctly gives the energy
for the transfer of an electron would also give a qualitatively correct
electron density and energy. For the set of HZ$^{\{2e\}}$ systems,
if the energy is wrong then the density will follow, giving rise to
a qualitatively incorrect transfer. An equivalent, but in our view, 
confusing way to phrase this challenge is about the Kohn-Sham potential.
To illustrate this point  take the case $Z=1.5$ for a stretched geometry
as an example. Let us ask the question what is the exact Kohn-Sham
potential that gives rise to integer number of electrons on each atom.
This is clear in the HZ$^{\{2e\}}$ system, as we have access to the
exact Kohn-Sham potential, $v_{s}(\mathbf{r})$ and $v_{xc}(\mathbf{r})$.
This comes from a very simple rearrangement of Eq. \ref{eq:KS equation},
substituting in the fact that the density is only made of one orbital, $\phi_{1}(\mathbf{r})=\sqrt{\rho(\mathbf{r})/2}$,
giving rise to
\begin{eqnarray}
v_{s}(\mathbf{r})+\epsilon_{1} & = & \frac{\nabla^{2}\rho}{4\rho}-\frac{(\nabla\rho)^{2}}{8\rho^{2}}\label{eq:vs in 2 electron}\\
v_{xc}(\mathbf{r})+\epsilon_{1} & = & \frac{\nabla^{2}\rho}{4\rho}-\frac{(\nabla\rho)^{2}}{8\rho^{2}}-v_{\mathrm{ext}}(\mathbf{r})-v_{J}(\mathbf{r}).\label{eq:vxc in 2 electron}
\end{eqnarray}
In this case, the exact Kohn-Sham potential is directly available
from an exact density in a FCI calculation. 

Many plots have been seen in the literature, in both ground state
and time-dependent analysis of the potential, for example \cite{perdew_what_1985,Gritsenko951870,Helbig09224105,tempel_revisiting_2009,Hellgren12022514}.
The way to understand any features of the potential is that they are
present as a consequence given a particular electron density, i.e.
the exact Kohn-Sham potential (Eq. \ref{eq:vs in 2 electron}) is
just a restatement of the density. This is shown in Fig. \ref{fig:Kohn-Sham-potential-HZ2e},
where we just evaluate Eq. \ref{eq:vs in 2 electron} with an already
minimised FCI density. If the potential shows any bumps it is because
it comes from a density that gives rise to that structure, not the
other way around. Furthermore, what gives rises to such a density
is what is energetically favourable (for example, to have one electron
each end), so it is the energy that is key. 

Fig. \ref{fig:Kohn-Sham-potential-HZ2e} is produced by FCI calculation
in a large basis set, but it is completely understandable and the
same as that given by a density of the form $\rho(\mathbf{r})=n_{1}e^{-2r}+n_{2}e^{-3(r-10)}$.
The divergences at the nuclei (labels (1)) are because $\nabla^{2}\rho$
goes as $\frac{1}{r}$. For an exponent $e^{-2\alpha r}$, the potential
at large $r$ goes to $v_{s}(\mathbf{r})+\epsilon_{1}=\alpha^{2}-\frac{\alpha^{2}}{2}$$ $,
so on the side of the H atom, e.g. label (2), it goes to a value of
$+\frac{1}{2}$ and on the side of the $Z=1.5$ atom it goes to a
value of $+\frac{9}{8}$ at label (3). The bump in the middle at label
(4) is where $\nabla\rho$ goes to zero because of the overlapping
densities, and the second term on the RHS of Eq. \ref{eq:vs in 2 electron}
disappears, so a value of roughly the average of $1^{2}$ and $1.5^{2}$
is obtained. Finally, the change at label (5) is because at long range
the $e^{-2r}$ from the H atom dominates over the $e^{-3r}$ behavior
from the Z atom, and from label (5) onwards the structure is a continuation
of the line approaching the bump at label (4). Understanding all these
features in the potential is of course relatively simple and known, and
it should help to dispel any mysterious nature of them. 

We want to stress that the bump in the middle (at label (4)) which
has repeatedly been related to the derivative discontinuity, is just
because $\nabla\rho$ goes to zero at some point in between the atoms
where the density is very small, and this can even be thought of as
a non-covalent interaction (NCI)\cite{Johnson106498}. It is not related
to any deeper physics and in particular is nothing to do with stopping
electrons moving from one side to another. It should also be noted
that the one thing that is not determined by this potential is the
constants in front of the density ($n_{1}$ and $n_{2}$), this always
cancels out. As such, it could be possible to have 0.8 electrons on
the H atom and 1.2 electrons on the Z atom with an identical potential
(note that the Z atom density would not respond to having more than
1 electron for the potential to remain unchanged). In general, these
steps and bumps do not stop the electrons moving, however, having
a correct energy functional does.

\section{Unrestricted Calculations}

Most of the problems we have highlighted to do with the integer nature
of electrons and the derivative discontinuity are in fact well captured
by unrestricted type methods. For example, the energies of infinitely
stretched H$_{2}^{+}$ and infinitely stretched H$_{2}$ are both
given by UHF. As H$_{2}$ is stretched, there is at some bond distance
a symmetry breaking (at the Coulson-Fisher point) beyond which the
alpha and beta densities can be found each on one of the atoms. Therefore,
the spin density is incorrect but the total density and energy are
very good. Also for HZ$^{\{2e\}}$ UHF recovers a discontinuity, and even
for the Hubbard model UHF does very well, giving qualitatively correct
gaps for all values of $U$.  Of course, there are other known problems
for UHF, such as stretching of F$_{2}$, for which RHF gives a reasonable
minimum in terms of geometry but the minimum is actually above the
dissociated atoms. RHF also dissociates incorrectly due to static
correlation error and in this case the change to UHF does more than
just correct the infinite limit, it actually means that UHF curve
has no minimum \cite{klimo_study_1981} (the Coulson-Fisher point
is before the minimum in the RHF curve). For F$_{2}$ one may argue
that a method such as UB3LYP gives a reasonable representation of
the stretching. Another similar example is given by the stretching
of O$_{2}^{2+}$\cite{Nobes1991216}.

\begin{figure}
\includegraphics[scale=0.33]{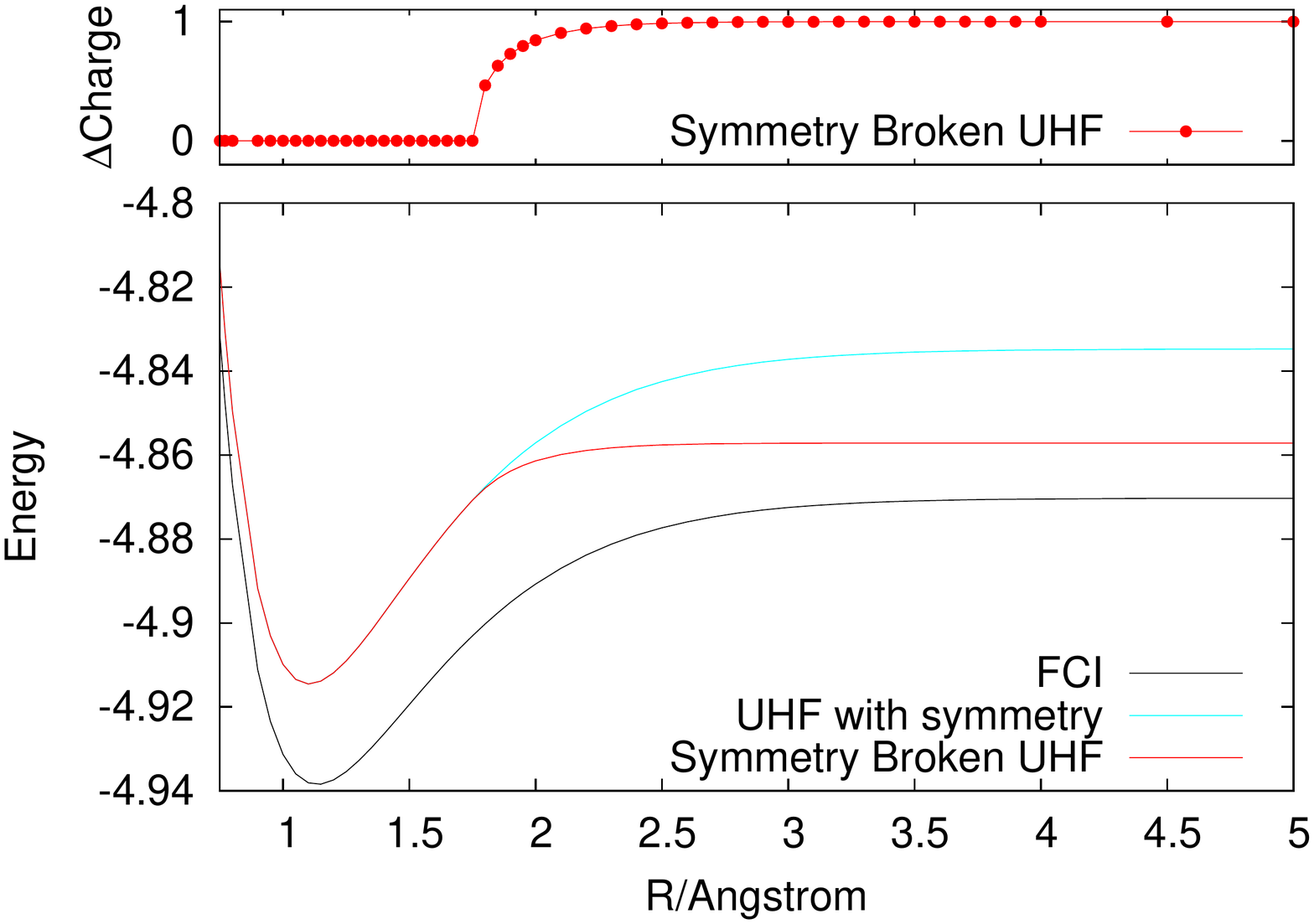} \caption{Unrestricted Hartree-Fock stretching of He$_{2}^{+}$ compared with
FCI. Shown at the top is the difference in charge between the two
He atoms. FCI has $\Delta{\rm Charge=0}$ for all bond distances,
however UHF incorrectly breaks the symmetry at around 1.8Å, with an
incorrect smooth transfer of half an electron from one atom to the
other. \label{fig:Unrestricted-Hartree-Fock-He2+}}
\end{figure}

Another system that shows up a qualitative problem of the UHF method
is that of stretching of He$_{2}^{+}$. Here, the problem is in symmetry
breaking, as illustrated in Fig. \ref{fig:Unrestricted-Hartree-Fock-He2+}.
Infinitely stretched He$_{2}^{+}$ with symmetry gives two He$^{\frac{1}{2}+}$
atoms.  UHF would give two high an energy due to its localization
error (a concave behaviour of the energy for fractional systems).
However, this too high energy is avoided by breaking the symmetry
to give dissociation products of He and He$^{+}$. This itself does
not seem wrong, but the examination of the dissociation curve indicates
that the symmetry breaking occurs at too short bond lengths, leading to
an incorrect smooth transfer of electrons around 1.8 Å. Also the UHF
curve falls off far too quickly compared to the FCI curve. This, of
course, is very similar to the symmetry breaking in the spin densities
seen in the UHF stretching of H$_{2},$ which is often argued to be
acceptable \cite{Perdew954531} as in H$_{2}$ the total density is
not qualitatively wrong, whereas in the case of He$_{2}^{+}$ it is
incorrect. For He$_{2}^{+},$ DFT methods such as UB3LYP dissociate
with half an electron on each atom but give a completely wrong and
much too low energy due to the delocalisation error. Similar symmetry
breaking by UHF can be seen in many systems, for example \cite{stein_stability_2014}.

\begin{figure*}
\subfloat[The HZ$^{1e}$ system with $Z=1.2$ for the wavefunctions $\Psi_{\alpha}=\sqrt{\left(\frac{1}{2}+\alpha\right)}\Psi_{(0)}+\sqrt{\left(\frac{1}{2}-\alpha\right)}\Psi_{(1)}$,
with $-\frac{1}{2}\le\alpha\le\frac{1}{2}$. PBE incorrectly gives
a minimum at $\alpha=0.24$, which corresponds to a qualitatively
incorrect wavefunction with fractional charges on both atoms.]{\includegraphics[scale=0.32]{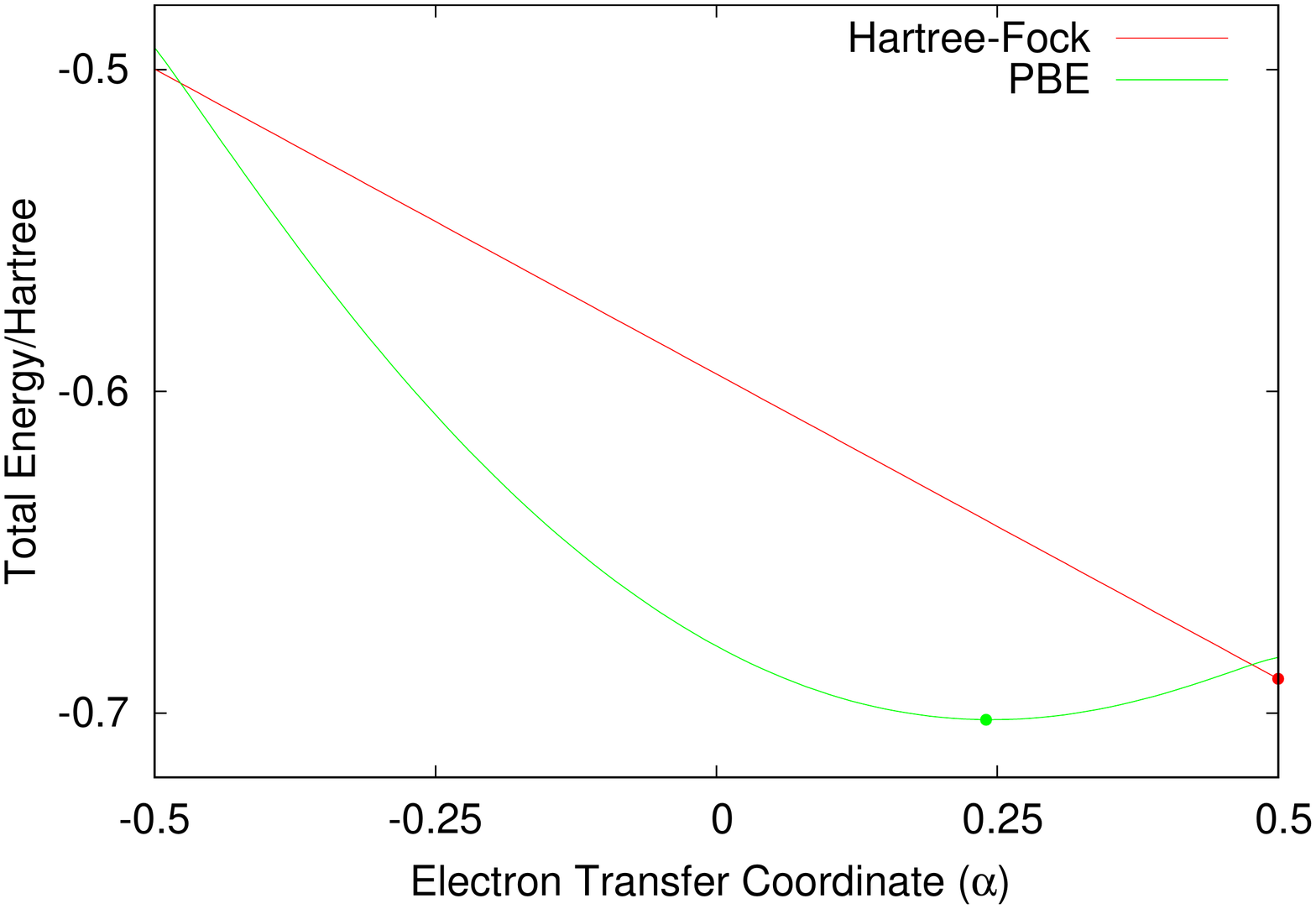}}\subfloat[HZ$^{\{2e\}}$ with $Z=1.2$ along the electron transfer coordinate
that moves both electrons from the H atom at $\alpha=-1$ to both
electrons on the Z atom at $\alpha=1$. For $\alpha>0$, $\Psi_{\alpha}=\sqrt{(1-\alpha)}\Psi_{(0)}+\sqrt{\alpha}\Psi_{(1)}$,
and $\alpha<0$ $\Psi_{\alpha}=\sqrt{(1-|\alpha|)}\Psi_{(0)}+\sqrt{|\alpha|}\Psi_{(2)}$.
PBE and RHF both transfer fractional numbers of electrons because
of the incorrect shape of their energy surfaces compared with the
FCI energy surface for electron transfer.]{\includegraphics[scale=0.32]{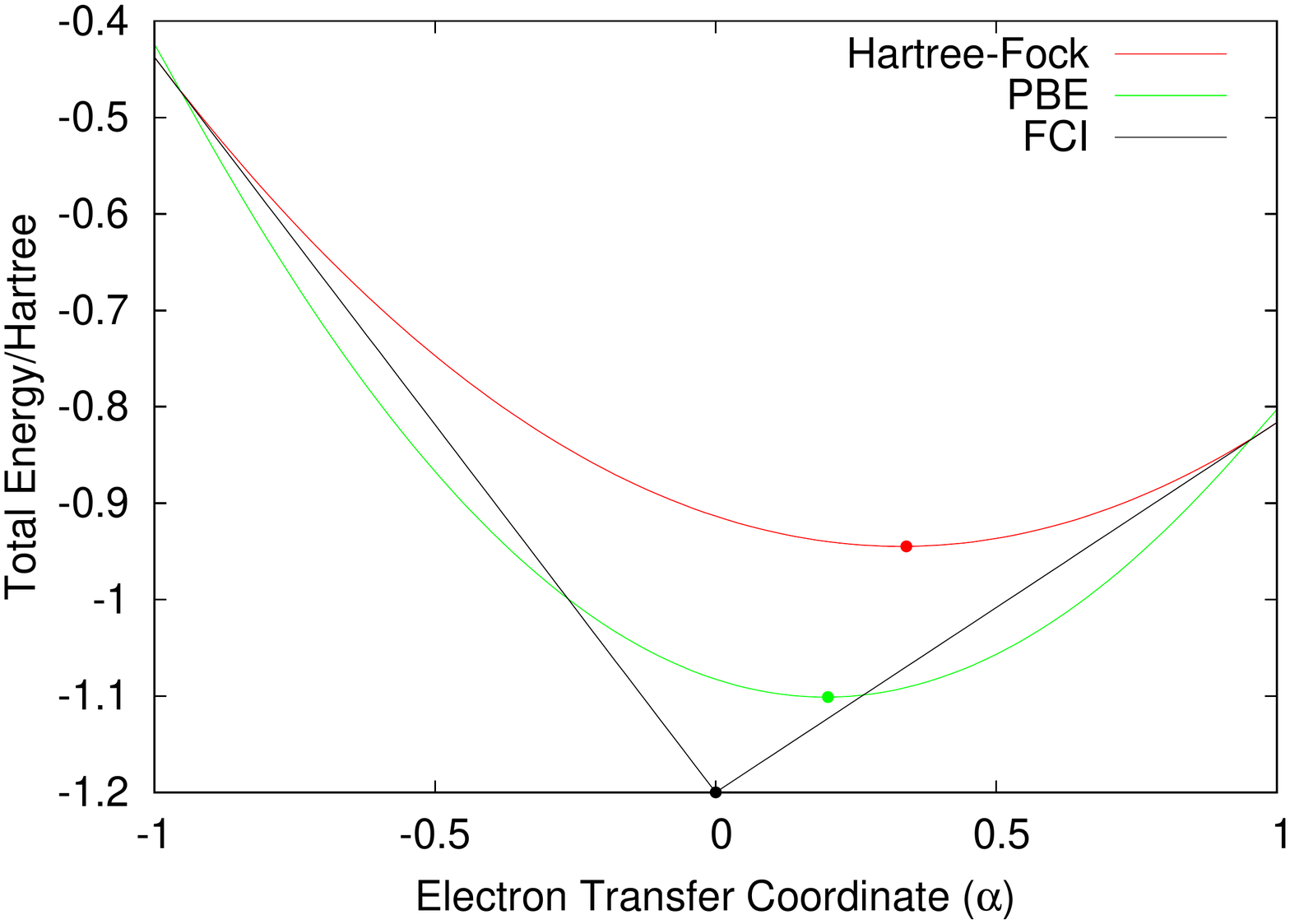}}\caption{\label{fig:Electron-transfer-lines}}
\end{figure*}

\begin{figure}[b]
\includegraphics[angle=-90,scale=0.37]{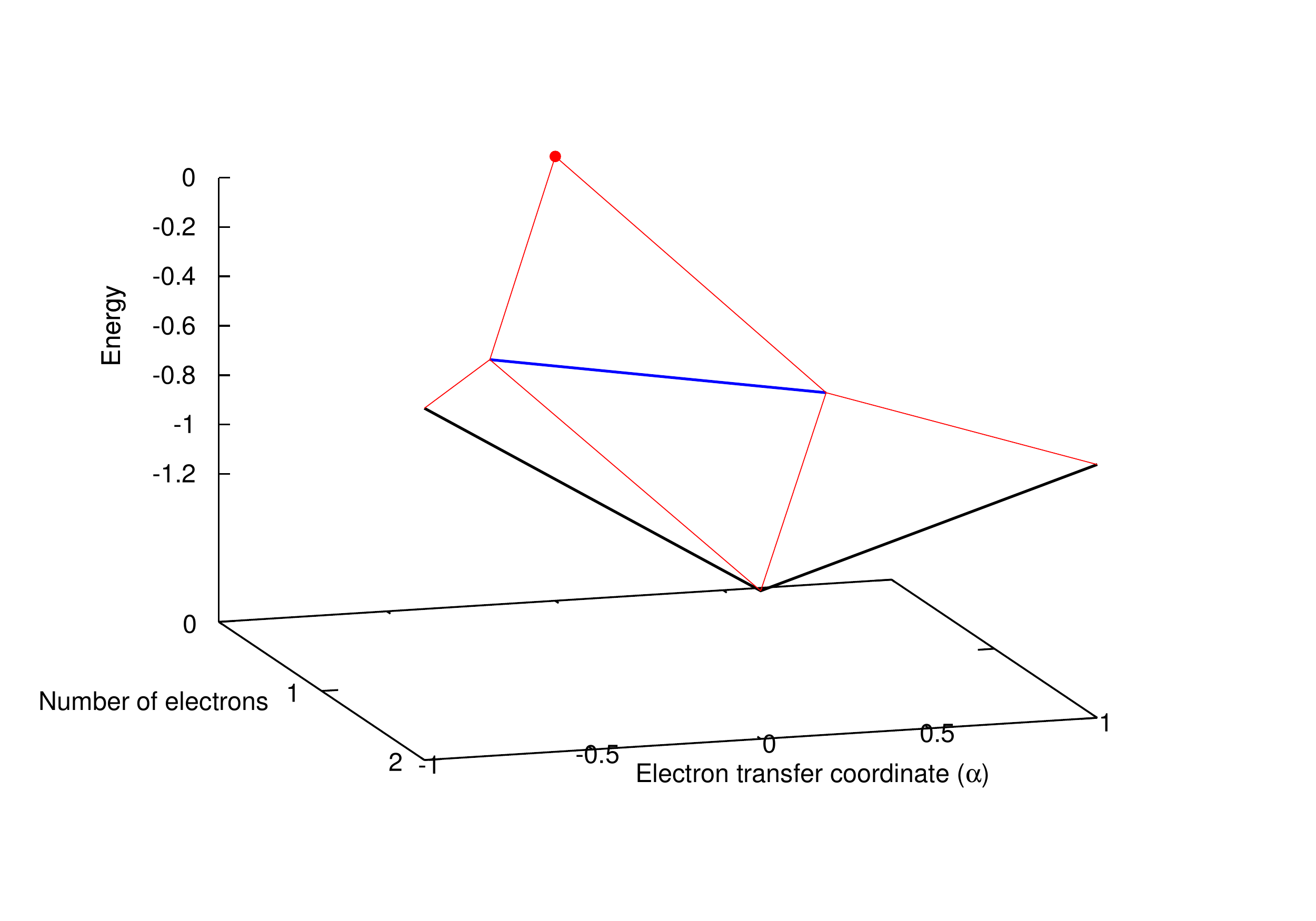}

\caption{The total energy of HZ with $Z=1.2$ along the coordinate of transferring
electrons for $0\le N\le2$. The line in blue shows the energy at
one electron and the line in black at two electrons, the red lines
correspond to fractional number of electrons and are given by PPLB.
The left hand side of the picture is when all the electron(s) are
on the H atom and the right-hand side when all the electron(s) are
on the Z atom.\label{fig:Frac transfer flat plane}}
\end{figure}

\section{Fractional electron transfer coordinate}

To understand in more detail the problem of the derivative discontinuity
in HZ, let us consider the simplest case of HZ stretched to large
distance (like 1000$a_{0}$). We now examine the transfer of electrons
from one end of the molecule to another, in just one particular HZ
system. In a single basis function per atom calculation this is very
easy, as FCI gives just a trivial number of states. Let us first look
at the system with one electron, where the FCI states are either one
electron on the H, $\Psi_{{\rm H}}$ with energy $E_{{\rm H}}$, or
one electron on the Z, $\Psi_{{\rm Z}}$ with energy $E_{{\rm Z}}$.
Let us now consider a state that is a general coherent sum of these,
$\Psi_{\alpha}=\sqrt{\alpha}\Psi_{{\rm H}}+\sqrt{(1-\alpha)}\Psi_{{\rm Z}}$,
$0\le\alpha\le1$. As $\Psi_{{\rm H}}$ and $\Psi_{{\rm Z}}$ are
FCI wavefunctions (one is the ground state the other is the first
excited state), they are orthogonal and eigenfunctions of the many-body
Hamiltonian, so it is trivially obtained that $E[\Psi_{\alpha}]=\alpha E_{{\rm H}}+(1-\alpha)E_{{\rm Z}}$
and $\rho_{\alpha}(\mathbf{r})=\alpha\rho_{{\rm H}}(\mathbf{r})+(1-\alpha)\rho_{{\rm Z}}(\mathbf{r})$.
This is very akin to fractional numbers of electrons as given PPLB
(Eq. \ref{eq:PPLBEnergy}), such that when $\alpha$ is varied the
energy varies linearly and the electron moves smoothly from the H
to the Z.  In contrast to PPLB all possible values of electron transfer,
$\alpha$, correspond to an integer system with one electron  and
are represented by a wavefunction.

HZ with two electrons can be analyzed similarly. In this case, at
the stretching limit, there are three singlet states with first order
density matrices $\left(\begin{array}{cc}
2 & 0\\
0 & 0
\end{array}\right)$, $\left(\begin{array}{cc}
1 & 0\\
0 & 1
\end{array}\right)$ and $\left(\begin{array}{cc}
0 & 0\\
0 & 2
\end{array}\right)$. The FCI wavefunctions that reduce down to give these density matrices
are eigenfunctions of the Hamiltonian and orthogonal, so a linear
combination of them will give a linear combination of both density
matrices and energies. With more basis functions the idea is similar
but the analysis more complex as there are many more possible states
(one basis function excludes any excited states of the atoms). 

Consider the case of HZ with $Z=1.2$ from FCI and compare it with
a functional such as PBE, as shown in Fig. \ref{fig:Electron-transfer-lines}.
First, for one electron (HZ$^{\{1e\}}$), where Hartree-Fock and FCI
are equivalent, there is a straight line interpolation between the
energies of the two atoms. Of course, a minimization of the FCI leads
to one electron wholly on one side or the other, in this case the
Z atom, as it is much lower in energy. The behavior of the energy
with a functional such as PBE is qualitatively incorrect for one electron,
as it does not have the correct linear straight line interpolation,
but instead the energy varies smoothly (almost parabolically) with
electron transfer. This leads to an incorrect minimum at around 0.26
electron on the H atom and 0.74 electron on the Z atom. The same result
could be found with the compatible fractional calculations on each
atom and piecing them back together, however, it is very good to see
it in an integer electron system with a corresponding wavefunction.
To summarise, in Fig. \ref{fig:Electron-transfer-lines}, PBE gives
a good energy for the ground state $\Psi=\Psi^{(0)}$ (and also for
the first excited state $\Psi=\Psi^{(1)})$ but incorrectly gives
a much lower energy for a wavefunction $\Psi=\sqrt{0.74}\Psi^{(0)}+\sqrt{0.26}\Psi^{(1)}$. 

For the two electron system (HZ$^{\{2e\}})$, FCI has a minimum with
one electron on each atom ($\Psi_{(0)})$ and two excited states,
one with two electrons on the Z atom ($\Psi_{(1)})$ and the other
with two electrons on the H atom ($\Psi_{(2)})$ . The straight line
interpolations to be considered are between the ground state and first
excited state and between the ground state and the second excited
state. These are the lowest energy ones, a higher energy would be
given by the interpolation between the first and second excited states.
For any of these wavefunctions, the HF and PBE energies can be trivially
evaluated from the density matrix, given that $T_{s}[\rho]$ for any
two electron system is the von-Weisacker expression$ $ $T_{s}^{{\rm vW}}[\rho]=\int\frac{(\nabla\rho({\bf r}))^{2}}{8\rho({\bf r})}\mathrm{d}{\bf r}$.
It is observed in Fig \ref{fig:Electron-transfer-lines}that both
HF and PBE qualitatively fail to describe the electron transfer in
this system. The energy on electron transfer is incorrectly smooth
for both methods and with minima at the wrong values, HF at 0.33 electron
on the H and 1.67 electron on the Z atom, and PBE with 0.4 electron
on the H and 1.6 electron on the Z atom. In terms of the wavefunction,
PBE is able to give a good energy for the excited states $\Psi=\Psi^{(1)}$
and $\Psi^{(2)}$, but due to its static correlation error PBE is
unable to give a good energy for $\Psi=\Psi^{(0)}$. This means that
PBE gives an incorrect minimum at $\Psi=\sqrt{0.4}\Psi^{(0)}+\sqrt{0.6}\Psi^{(1)}$.
In general, density functional approximations favour an incorrect
fractional charge transfer. Understanding all the possible states
of this one systems is perhaps a much simpler challenge than understanding
many different system (i.e. changing the molecule by changing $Z$)
and yet captures the same physics.

\section{Perspectives for the future}

The key picture of the derivative discontinuity of the total energy
is shown by FCI calculations in Figs \ref{fig:H8 1basis} and \ref{fig:H8 closed shell line },
where the density increases smoothly but the energy is discontinuous
on passing through one electron per site. Therefore, there is an intrinsic
discontinuity in the exchange-correlation term that is a consequence
of the particle nature of electrons. Approximate functionals in the
literature completely miss this behaviour and the failure in the total
energies is clear as, for example, in the qualitative breakdown to
describe the energies of stretched H$_{2}$ and stretched H$_{2}^{+}$.
This error can be transformed into an error in the density as shown
in the two electron example, HZ$^{\{2e\}}$ (Fig. \ref{fig:Occupation-HZ2e}).
This leads to qualitative failures to describe charge transfer, with
an artificial bias to fractionally transfer electrons. Remarkably,
this is seen in systems with integer number of electrons characterized
by a wavefunction, in contrast to the delocalization error typical
of fractionally charged systems.

It is clear that the problems caused by missing the derivative discontinuity
are not just about stretching molecules, but it is the same physics
that occurs in transition metal complexes, chemical reactions, or
especially in electron transfer processes. Our hope is that the use
of simple chemical model systems gives a more complete understanding
of the nature of all types of electronic behaviour that occur in the
intrincate nature of electronic structure. 

The bumps of the exchange-correlation potential at the bond regions,
where the density is very small, have been shown to have no effect
on how electrons move and, therefore, do not capture the challenge
of the derivative discontinuity. This illustrates the point that when
the understanding appears quite paradoxical, it is just a clue that
a deeper comprehension of the problem is needed. 

We have highlighted the importance of the derivative discontinuity
as a challenge for the energy functional at an integer number of electrons.
We hope that understanding the examples given in this work can help
highlight the avenue for development of new exchange-correlation functionals
that contain the physics of the derivative discontinuity and represent
the integer nature of electrons. This is needed so that DFT can play
its role in helping tackle many important technological applications
by correctly describing the movements of electrons in systems such
as batteries and solar cells, chemical reactions in proteins, and
transition metal compounds.

\begin{acknowledgments}
We gratefully acknowledge funding from the Royal Society (AJC) and
Ramon y Cajal (PMS). PMS also acknowledges grant FIS2012-37549 from
the Spanish Ministry of Science.
\end{acknowledgments}
\bibliographystyle{apsrev4-1}

\begin{thebibliography}{80}%
\makeatletter
\providecommand \@ifxundefined [1]{%
 \@ifx{#1\undefined}
}%
\providecommand \@ifnum [1]{%
 \ifnum #1\expandafter \@firstoftwo
 \else \expandafter \@secondoftwo
 \fi
}%
\providecommand \@ifx [1]{%
 \ifx #1\expandafter \@firstoftwo
 \else \expandafter \@secondoftwo
 \fi
}%
\providecommand \natexlab [1]{#1}%
\providecommand \enquote  [1]{``#1''}%
\providecommand \bibnamefont  [1]{#1}%
\providecommand \bibfnamefont [1]{#1}%
\providecommand \citenamefont [1]{#1}%
\providecommand \href@noop [0]{\@secondoftwo}%
\providecommand \href [0]{\begingroup \@sanitize@url \@href}%
\providecommand \@href[1]{\@@startlink{#1}\@@href}%
\providecommand \@@href[1]{\endgroup#1\@@endlink}%
\providecommand \@sanitize@url [0]{\catcode `\\12\catcode `\$12\catcode
  `\&12\catcode `\#12\catcode `\^12\catcode `\_12\catcode `\%12\relax}%
\providecommand \@@startlink[1]{}%
\providecommand \@@endlink[0]{}%
\providecommand \url  [0]{\begingroup\@sanitize@url \@url }%
\providecommand \@url [1]{\endgroup\@href {#1}{\urlprefix }}%
\providecommand \urlprefix  [0]{URL }%
\providecommand \Eprint [0]{\href }%
\providecommand \doibase [0]{http://dx.doi.org/}%
\providecommand \selectlanguage [0]{\@gobble}%
\providecommand \bibinfo  [0]{\@secondoftwo}%
\providecommand \bibfield  [0]{\@secondoftwo}%
\providecommand \translation [1]{[#1]}%
\providecommand \BibitemOpen [0]{}%
\providecommand \bibitemStop [0]{}%
\providecommand \bibitemNoStop [0]{.\EOS\space}%
\providecommand \EOS [0]{\spacefactor3000\relax}%
\providecommand \BibitemShut  [1]{\csname bibitem#1\endcsname}%
\let\auto@bib@innerbib\@empty
\bibitem [{\citenamefont {Dirac}(1930)}]{Dirac30}%
  \BibitemOpen
  \bibfield  {author} {\bibinfo {author} {\bibfnamefont {P.~A.~M.}\
  \bibnamefont {Dirac}},\ }\href@noop {} {\bibfield  {journal} {\bibinfo
  {journal} {Proc. Cambridge Philos. Soc.}\ }\textbf {\bibinfo {volume} {26}},\
  \bibinfo {pages} {376} (\bibinfo {year} {1930})}\BibitemShut {NoStop}%
\bibitem [{\citenamefont {Vosko}\ \emph {et~al.}(1980)\citenamefont {Vosko},
  \citenamefont {Wilk},\ and\ \citenamefont {Nusair}}]{Vosko801200}%
  \BibitemOpen
  \bibfield  {author} {\bibinfo {author} {\bibfnamefont {S.~H.}\ \bibnamefont
  {Vosko}}, \bibinfo {author} {\bibfnamefont {L.}~\bibnamefont {Wilk}}, \ and\
  \bibinfo {author} {\bibfnamefont {M.}~\bibnamefont {Nusair}},\ }\href@noop {}
  {\bibfield  {journal} {\bibinfo  {journal} {Can. J. Phys.}\ }\textbf
  {\bibinfo {volume} {58}},\ \bibinfo {pages} {1200} (\bibinfo {year}
  {1980})}\BibitemShut {NoStop}%
\bibitem [{\citenamefont {Perdew}\ and\ \citenamefont
  {Wang}(1992)}]{Perdew9213244}%
  \BibitemOpen
  \bibfield  {author} {\bibinfo {author} {\bibfnamefont {J.~P.}\ \bibnamefont
  {Perdew}}\ and\ \bibinfo {author} {\bibfnamefont {Y.}~\bibnamefont {Wang}},\
  }\href@noop {} {\bibfield  {journal} {\bibinfo  {journal} {Phys. Rev. B}\
  }\textbf {\bibinfo {volume} {45}},\ \bibinfo {pages} {13244} (\bibinfo {year}
  {1992})}\BibitemShut {NoStop}%
\bibitem [{\citenamefont {Becke}(1988{\natexlab{a}})}]{Becke883098}%
  \BibitemOpen
  \bibfield  {author} {\bibinfo {author} {\bibfnamefont {A.~D.}\ \bibnamefont
  {Becke}},\ }\href@noop {} {\bibfield  {journal} {\bibinfo  {journal} {Phys.
  Rev. A}\ }\textbf {\bibinfo {volume} {38}},\ \bibinfo {pages} {3098}
  (\bibinfo {year} {1988}{\natexlab{a}})}\BibitemShut {NoStop}%
\bibitem [{\citenamefont {Lee}\ \emph {et~al.}(1988)\citenamefont {Lee},
  \citenamefont {Yang},\ and\ \citenamefont {Parr}}]{Lee88785}%
  \BibitemOpen
  \bibfield  {author} {\bibinfo {author} {\bibfnamefont {C.~T.}\ \bibnamefont
  {Lee}}, \bibinfo {author} {\bibfnamefont {W.~T.}\ \bibnamefont {Yang}}, \
  and\ \bibinfo {author} {\bibfnamefont {R.~G.}\ \bibnamefont {Parr}},\
  }\href@noop {} {\bibfield  {journal} {\bibinfo  {journal} {Phys. Rev. B}\
  }\textbf {\bibinfo {volume} {37}},\ \bibinfo {pages} {785} (\bibinfo {year}
  {1988})}\BibitemShut {NoStop}%
\bibitem [{\citenamefont {Perdew}\ \emph {et~al.}(1996)\citenamefont {Perdew},
  \citenamefont {Burke},\ and\ \citenamefont {Ernzerhof}}]{Perdew963865}%
  \BibitemOpen
  \bibfield  {author} {\bibinfo {author} {\bibfnamefont {J.~P.}\ \bibnamefont
  {Perdew}}, \bibinfo {author} {\bibfnamefont {K.}~\bibnamefont {Burke}}, \
  and\ \bibinfo {author} {\bibfnamefont {M.}~\bibnamefont {Ernzerhof}},\
  }\href@noop {} {\bibfield  {journal} {\bibinfo  {journal} {Phys. Rev. Lett.}\
  }\textbf {\bibinfo {volume} {77}},\ \bibinfo {pages} {3865} (\bibinfo {year}
  {1996})}\BibitemShut {NoStop}%
\bibitem [{\citenamefont {Perdew}(1986)}]{Perdew868822}%
  \BibitemOpen
  \bibfield  {author} {\bibinfo {author} {\bibfnamefont {J.~P.}\ \bibnamefont
  {Perdew}},\ }\href@noop {} {\bibfield  {journal} {\bibinfo  {journal} {Phys.
  Rev. B}\ }\textbf {\bibinfo {volume} {33}},\ \bibinfo {pages} {8822}
  (\bibinfo {year} {1986})}\BibitemShut {NoStop}%
\bibitem [{\citenamefont {Handy}\ and\ \citenamefont
  {Cohen}(2001)}]{Handy01403}%
  \BibitemOpen
  \bibfield  {author} {\bibinfo {author} {\bibfnamefont {N.~C.}\ \bibnamefont
  {Handy}}\ and\ \bibinfo {author} {\bibfnamefont {A.~J.}\ \bibnamefont
  {Cohen}},\ }\href@noop {} {\bibfield  {journal} {\bibinfo  {journal} {Mol.
  Phys.}\ }\textbf {\bibinfo {volume} {99}},\ \bibinfo {pages} {403} (\bibinfo
  {year} {2001})}\BibitemShut {NoStop}%
\bibitem [{\citenamefont {Becke}(1988{\natexlab{b}})}]{Becke881053}%
  \BibitemOpen
  \bibfield  {author} {\bibinfo {author} {\bibfnamefont {A.~D.}\ \bibnamefont
  {Becke}},\ }\href@noop {} {\bibfield  {journal} {\bibinfo  {journal} {J.
  Chem. Phys.}\ }\textbf {\bibinfo {volume} {88}},\ \bibinfo {pages} {1053}
  (\bibinfo {year} {1988}{\natexlab{b}})}\BibitemShut {NoStop}%
\bibitem [{\citenamefont {Becke}\ and\ \citenamefont
  {Roussel}(1989)}]{Becke893761}%
  \BibitemOpen
  \bibfield  {author} {\bibinfo {author} {\bibfnamefont {A.~D.}\ \bibnamefont
  {Becke}}\ and\ \bibinfo {author} {\bibfnamefont {M.~R.}\ \bibnamefont
  {Roussel}},\ }\href@noop {} {\bibfield  {journal} {\bibinfo  {journal} {Phys.
  Rev. A}\ }\textbf {\bibinfo {volume} {39}},\ \bibinfo {pages} {3761}
  (\bibinfo {year} {1989})}\BibitemShut {NoStop}%
\bibitem [{\citenamefont {Tao}\ \emph {et~al.}(2003)\citenamefont {Tao},
  \citenamefont {Perdew}, \citenamefont {Staroverov},\ and\ \citenamefont
  {Scuseria}}]{Tao03146401}%
  \BibitemOpen
  \bibfield  {author} {\bibinfo {author} {\bibfnamefont {J.~M.}\ \bibnamefont
  {Tao}}, \bibinfo {author} {\bibfnamefont {J.~P.}\ \bibnamefont {Perdew}},
  \bibinfo {author} {\bibfnamefont {V.~N.}\ \bibnamefont {Staroverov}}, \ and\
  \bibinfo {author} {\bibfnamefont {G.~E.}\ \bibnamefont {Scuseria}},\
  }\href@noop {} {\bibfield  {journal} {\bibinfo  {journal} {Phys. Rev. Lett.}\
  }\textbf {\bibinfo {volume} {91}},\ \bibinfo {pages} {146401} (\bibinfo
  {year} {2003})}\BibitemShut {NoStop}%
\bibitem [{\citenamefont {Zhao}\ and\ \citenamefont
  {Truhlar}(2006)}]{Zhao06194101}%
  \BibitemOpen
  \bibfield  {author} {\bibinfo {author} {\bibfnamefont {Y.}~\bibnamefont
  {Zhao}}\ and\ \bibinfo {author} {\bibfnamefont {D.~G.}\ \bibnamefont
  {Truhlar}},\ }\href@noop {} {\bibfield  {journal} {\bibinfo  {journal} {J.
  Chem. Phys.}\ }\textbf {\bibinfo {volume} {125}},\ \bibinfo {pages} {194101}
  (\bibinfo {year} {2006})}\BibitemShut {NoStop}%
\bibitem [{\citenamefont {Becke}(1993{\natexlab{a}})}]{Becke931372}%
  \BibitemOpen
  \bibfield  {author} {\bibinfo {author} {\bibfnamefont {A.~D.}\ \bibnamefont
  {Becke}},\ }\href@noop {} {\bibfield  {journal} {\bibinfo  {journal} {J.
  Chem. Phys.}\ }\textbf {\bibinfo {volume} {98}},\ \bibinfo {pages} {1372}
  (\bibinfo {year} {1993}{\natexlab{a}})}\BibitemShut {NoStop}%
\bibitem [{\citenamefont {Becke}(1993{\natexlab{b}})}]{Becke935648}%
  \BibitemOpen
  \bibfield  {author} {\bibinfo {author} {\bibfnamefont {A.~D.}\ \bibnamefont
  {Becke}},\ }\href@noop {} {\bibfield  {journal} {\bibinfo  {journal} {J.
  Chem. Phys.}\ }\textbf {\bibinfo {volume} {98}},\ \bibinfo {pages} {5648}
  (\bibinfo {year} {1993}{\natexlab{b}})}\BibitemShut {NoStop}%
\bibitem [{\citenamefont {Stephens}\ \emph {et~al.}(1994)\citenamefont
  {Stephens}, \citenamefont {Devlin}, \citenamefont {Chabalowski},\ and\
  \citenamefont {Frisch}}]{Stephens9411623}%
  \BibitemOpen
  \bibfield  {author} {\bibinfo {author} {\bibfnamefont {P.~J.}\ \bibnamefont
  {Stephens}}, \bibinfo {author} {\bibfnamefont {F.~J.}\ \bibnamefont
  {Devlin}}, \bibinfo {author} {\bibfnamefont {C.~F.}\ \bibnamefont
  {Chabalowski}}, \ and\ \bibinfo {author} {\bibfnamefont {M.~J.}\ \bibnamefont
  {Frisch}},\ }\href@noop {} {\bibfield  {journal} {\bibinfo  {journal} {J.
  Phys. Chem.}\ }\textbf {\bibinfo {volume} {98}},\ \bibinfo {pages} {11623}
  (\bibinfo {year} {1994})}\BibitemShut {NoStop}%
\bibitem [{\citenamefont {Adamo}\ and\ \citenamefont
  {Barone}(1999)}]{Adamo996158}%
  \BibitemOpen
  \bibfield  {author} {\bibinfo {author} {\bibfnamefont {C.}~\bibnamefont
  {Adamo}}\ and\ \bibinfo {author} {\bibfnamefont {V.}~\bibnamefont {Barone}},\
  }\href@noop {} {\bibfield  {journal} {\bibinfo  {journal} {J. Chem. Phys.}\
  }\textbf {\bibinfo {volume} {110}},\ \bibinfo {pages} {6158} (\bibinfo {year}
  {1999})}\BibitemShut {NoStop}%
\bibitem [{\citenamefont {Becke}(1997)}]{Becke978554}%
  \BibitemOpen
  \bibfield  {author} {\bibinfo {author} {\bibfnamefont {A.~D.}\ \bibnamefont
  {Becke}},\ }\href@noop {} {\bibfield  {journal} {\bibinfo  {journal} {J.
  Chem. Phys.}\ }\textbf {\bibinfo {volume} {107}},\ \bibinfo {pages} {8554}
  (\bibinfo {year} {1997})}\BibitemShut {NoStop}%
\bibitem [{\citenamefont {Hamprecht}\ \emph {et~al.}(1998)\citenamefont
  {Hamprecht}, \citenamefont {Cohen}, \citenamefont {Tozer},\ and\
  \citenamefont {Handy}}]{Hamprecht986264}%
  \BibitemOpen
  \bibfield  {author} {\bibinfo {author} {\bibfnamefont {F.~A.}\ \bibnamefont
  {Hamprecht}}, \bibinfo {author} {\bibfnamefont {A.~J.}\ \bibnamefont
  {Cohen}}, \bibinfo {author} {\bibfnamefont {D.~J.}\ \bibnamefont {Tozer}}, \
  and\ \bibinfo {author} {\bibfnamefont {N.~C.}\ \bibnamefont {Handy}},\
  }\href@noop {} {\bibfield  {journal} {\bibinfo  {journal} {J. Chem. Phys.}\
  }\textbf {\bibinfo {volume} {109}},\ \bibinfo {pages} {6264} (\bibinfo {year}
  {1998})}\BibitemShut {NoStop}%
\bibitem [{\citenamefont {Keal}\ and\ \citenamefont
  {Tozer}(2005)}]{Keal05121103}%
  \BibitemOpen
  \bibfield  {author} {\bibinfo {author} {\bibfnamefont {T.~W.}\ \bibnamefont
  {Keal}}\ and\ \bibinfo {author} {\bibfnamefont {D.~J.}\ \bibnamefont
  {Tozer}},\ }\href@noop {} {\bibfield  {journal} {\bibinfo  {journal} {J.
  Chem. Phys.}\ }\textbf {\bibinfo {volume} {123}},\ \bibinfo {pages} {121103}
  (\bibinfo {year} {2005})}\BibitemShut {NoStop}%
\bibitem [{\citenamefont {Iikura}\ \emph {et~al.}(2001)\citenamefont {Iikura},
  \citenamefont {Tsuneda}, \citenamefont {Yanai},\ and\ \citenamefont
  {Hirao}}]{Iikura013540}%
  \BibitemOpen
  \bibfield  {author} {\bibinfo {author} {\bibfnamefont {H.}~\bibnamefont
  {Iikura}}, \bibinfo {author} {\bibfnamefont {T.}~\bibnamefont {Tsuneda}},
  \bibinfo {author} {\bibfnamefont {T.}~\bibnamefont {Yanai}}, \ and\ \bibinfo
  {author} {\bibfnamefont {K.}~\bibnamefont {Hirao}},\ }\href@noop {}
  {\bibfield  {journal} {\bibinfo  {journal} {J. Chem. Phys.}\ }\textbf
  {\bibinfo {volume} {115}},\ \bibinfo {pages} {3540} (\bibinfo {year}
  {2001})}\BibitemShut {NoStop}%
\bibitem [{\citenamefont {Yanai}\ \emph {et~al.}(2004)\citenamefont {Yanai},
  \citenamefont {Tew},\ and\ \citenamefont {Handy}}]{Yanai0451}%
  \BibitemOpen
  \bibfield  {author} {\bibinfo {author} {\bibfnamefont {T.}~\bibnamefont
  {Yanai}}, \bibinfo {author} {\bibfnamefont {D.~P.}\ \bibnamefont {Tew}}, \
  and\ \bibinfo {author} {\bibfnamefont {N.~C.}\ \bibnamefont {Handy}},\
  }\href@noop {} {\bibfield  {journal} {\bibinfo  {journal} {Chem. Phys.
  Lett.}\ }\textbf {\bibinfo {volume} {393}},\ \bibinfo {pages} {51} (\bibinfo
  {year} {2004})}\BibitemShut {NoStop}%
\bibitem [{\citenamefont {Vydrov}\ and\ \citenamefont
  {Scuseria}(2006)}]{Vydrov06234109}%
  \BibitemOpen
  \bibfield  {author} {\bibinfo {author} {\bibfnamefont {O.~A.}\ \bibnamefont
  {Vydrov}}\ and\ \bibinfo {author} {\bibfnamefont {G.~E.}\ \bibnamefont
  {Scuseria}},\ }\href@noop {} {\bibfield  {journal} {\bibinfo  {journal} {J.
  Chem. Phys.}\ }\textbf {\bibinfo {volume} {125}},\ \bibinfo {pages} {234109}
  (\bibinfo {year} {2006})}\BibitemShut {NoStop}%
\bibitem [{\citenamefont {Chai}\ and\ \citenamefont
  {Head-Gordon}(2008)}]{Chai086615}%
  \BibitemOpen
  \bibfield  {author} {\bibinfo {author} {\bibfnamefont {J.~D.}\ \bibnamefont
  {Chai}}\ and\ \bibinfo {author} {\bibfnamefont {M.}~\bibnamefont
  {Head-Gordon}},\ }\href@noop {} {\bibfield  {journal} {\bibinfo  {journal}
  {Phys. Chem. Chem. Phys.}\ }\textbf {\bibinfo {volume} {10}},\ \bibinfo
  {pages} {6615} (\bibinfo {year} {2008})}\BibitemShut {NoStop}%
\bibitem [{\citenamefont {Heyd}\ \emph {et~al.}(2003)\citenamefont {Heyd},
  \citenamefont {Scuseria},\ and\ \citenamefont {Ernzerhof}}]{Heyd038207}%
  \BibitemOpen
  \bibfield  {author} {\bibinfo {author} {\bibfnamefont {J.}~\bibnamefont
  {Heyd}}, \bibinfo {author} {\bibfnamefont {G.~E.}\ \bibnamefont {Scuseria}},
  \ and\ \bibinfo {author} {\bibfnamefont {M.}~\bibnamefont {Ernzerhof}},\
  }\href@noop {} {\bibfield  {journal} {\bibinfo  {journal} {J. Chem. Phys.}\
  }\textbf {\bibinfo {volume} {118}},\ \bibinfo {pages} {8207} (\bibinfo {year}
  {2003})}\BibitemShut {NoStop}%
\bibitem [{\citenamefont {Heyd}\ \emph {et~al.}(2006)\citenamefont {Heyd},
  \citenamefont {Scuseria},\ and\ \citenamefont {Ernzerhof}}]{Heyd06219906}%
  \BibitemOpen
  \bibfield  {author} {\bibinfo {author} {\bibfnamefont {J.}~\bibnamefont
  {Heyd}}, \bibinfo {author} {\bibfnamefont {G.~E.}\ \bibnamefont {Scuseria}},
  \ and\ \bibinfo {author} {\bibfnamefont {M.}~\bibnamefont {Ernzerhof}},\
  }\href@noop {} {\bibfield  {journal} {\bibinfo  {journal} {J. Chem. Phys.}\
  }\textbf {\bibinfo {volume} {124}},\ \bibinfo {pages} {219906} (\bibinfo
  {year} {2006})}\BibitemShut {NoStop}%
\bibitem [{\citenamefont {Perdew}\ and\ \citenamefont
  {Schmidt}(2001)}]{Perdew011}%
  \BibitemOpen
  \bibfield  {author} {\bibinfo {author} {\bibfnamefont {J.~P.}\ \bibnamefont
  {Perdew}}\ and\ \bibinfo {author} {\bibfnamefont {K.}~\bibnamefont
  {Schmidt}},\ }in\ \href@noop {} {\emph {\bibinfo {booktitle} {Density
  Functional Theory and Its Application to Materials}}},\ \bibinfo {series}
  {Aip Conference Proceedings}, Vol.\ \bibinfo {volume} {577},\ \bibinfo
  {editor} {edited by\ \bibinfo {editor} {\bibfnamefont {V.}~\bibnamefont
  {VanDoren}}, \bibinfo {editor} {\bibfnamefont {C.}~\bibnamefont
  {VanAlsenoy}}, \ and\ \bibinfo {editor} {\bibfnamefont {P.}~\bibnamefont
  {Geerlings}}}\ (\bibinfo {year} {2001})\ pp.\ \bibinfo {pages}
  {1--20}\BibitemShut {NoStop}%
\bibitem [{\citenamefont {Grimme}(2006)}]{Grimme0634108}%
  \BibitemOpen
  \bibfield  {author} {\bibinfo {author} {\bibfnamefont {S.}~\bibnamefont
  {Grimme}},\ }\href@noop {} {\bibfield  {journal} {\bibinfo  {journal} {J.
  Chem. Phys.}\ }\textbf {\bibinfo {volume} {124}},\ \bibinfo {pages} {034108}
  (\bibinfo {year} {2006})}\BibitemShut {NoStop}%
\bibitem [{\citenamefont {Bohm}\ and\ \citenamefont {Pines}(1952)}]{RPA1}%
  \BibitemOpen
  \bibfield  {author} {\bibinfo {author} {\bibfnamefont {D.}~\bibnamefont
  {Bohm}}\ and\ \bibinfo {author} {\bibfnamefont {D.}~\bibnamefont {Pines}},\
  }\href@noop {} {\bibfield  {journal} {\bibinfo  {journal} {Phys. Rev.}\
  }\textbf {\bibinfo {volume} {85}},\ \bibinfo {pages} {332} (\bibinfo {year}
  {1952})}\BibitemShut {NoStop}%
\bibitem [{\citenamefont {Furche}(2008)}]{Furche08114105}%
  \BibitemOpen
  \bibfield  {author} {\bibinfo {author} {\bibfnamefont {F.}~\bibnamefont
  {Furche}},\ }\href@noop {} {\bibfield  {journal} {\bibinfo  {journal} {J.
  Chem. Phys.}\ }\textbf {\bibinfo {volume} {129}},\ \bibinfo {pages} {114105}
  (\bibinfo {year} {2008})}\BibitemShut {NoStop}%
\bibitem [{\citenamefont {Cohen}\ \emph {et~al.}(2012)\citenamefont {Cohen},
  \citenamefont {Mori-S\'anchez},\ and\ \citenamefont {Yang}}]{Cohen12289}%
  \BibitemOpen
  \bibfield  {author} {\bibinfo {author} {\bibfnamefont {A.~J.}\ \bibnamefont
  {Cohen}}, \bibinfo {author} {\bibfnamefont {P.}~\bibnamefont
  {Mori-S\'anchez}}, \ and\ \bibinfo {author} {\bibfnamefont {W.~T.}\
  \bibnamefont {Yang}},\ }\href@noop {} {\bibfield  {journal} {\bibinfo
  {journal} {Chem. Rev.}\ }\textbf {\bibinfo {volume} {112}},\ \bibinfo {pages}
  {289} (\bibinfo {year} {2012})}\BibitemShut {NoStop}%
\bibitem [{\citenamefont {Curtiss}\ \emph {et~al.}(2000)\citenamefont
  {Curtiss}, \citenamefont {Raghavachari}, \citenamefont {Redfern},\ and\
  \citenamefont {Pople}}]{Curtiss007374}%
  \BibitemOpen
  \bibfield  {author} {\bibinfo {author} {\bibfnamefont {L.~A.}\ \bibnamefont
  {Curtiss}}, \bibinfo {author} {\bibfnamefont {K.}~\bibnamefont
  {Raghavachari}}, \bibinfo {author} {\bibfnamefont {P.~C.}\ \bibnamefont
  {Redfern}}, \ and\ \bibinfo {author} {\bibfnamefont {J.~A.}\ \bibnamefont
  {Pople}},\ }\href@noop {} {\bibfield  {journal} {\bibinfo  {journal} {J.
  Chem. Phys.}\ }\textbf {\bibinfo {volume} {112}},\ \bibinfo {pages} {7374}
  (\bibinfo {year} {2000})}\BibitemShut {NoStop}%
\bibitem [{\citenamefont {Zhao}\ \emph {et~al.}(2004)\citenamefont {Zhao},
  \citenamefont {Lynch},\ and\ \citenamefont {Truhlar}}]{Zhao042715}%
  \BibitemOpen
  \bibfield  {author} {\bibinfo {author} {\bibfnamefont {Y.}~\bibnamefont
  {Zhao}}, \bibinfo {author} {\bibfnamefont {B.~J.}\ \bibnamefont {Lynch}}, \
  and\ \bibinfo {author} {\bibfnamefont {D.~G.}\ \bibnamefont {Truhlar}},\
  }\href@noop {} {\bibfield  {journal} {\bibinfo  {journal} {J. Phys. Chem. A}\
  }\textbf {\bibinfo {volume} {108}},\ \bibinfo {pages} {2715} (\bibinfo {year}
  {2004})}\BibitemShut {NoStop}%
\bibitem [{\citenamefont {Zhao}\ \emph {et~al.}(2005)\citenamefont {Zhao},
  \citenamefont {Gonzalez-Garcia},\ and\ \citenamefont {Truhlar}}]{Zhao052012}%
  \BibitemOpen
  \bibfield  {author} {\bibinfo {author} {\bibfnamefont {Y.}~\bibnamefont
  {Zhao}}, \bibinfo {author} {\bibfnamefont {N.}~\bibnamefont
  {Gonzalez-Garcia}}, \ and\ \bibinfo {author} {\bibfnamefont {D.~G.}\
  \bibnamefont {Truhlar}},\ }\href@noop {} {\bibfield  {journal} {\bibinfo
  {journal} {J. Phys. Chem. A}\ }\textbf {\bibinfo {volume} {109}},\ \bibinfo
  {pages} {2012} (\bibinfo {year} {2005})}\BibitemShut {NoStop}%
\bibitem [{\citenamefont {Curtiss}\ \emph {et~al.}(1997)\citenamefont
  {Curtiss}, \citenamefont {Raghavachari}, \citenamefont {Redfern},\ and\
  \citenamefont {Pople}}]{Curtiss971063}%
  \BibitemOpen
  \bibfield  {author} {\bibinfo {author} {\bibfnamefont {L.}~\bibnamefont
  {Curtiss}}, \bibinfo {author} {\bibfnamefont {K.}~\bibnamefont
  {Raghavachari}}, \bibinfo {author} {\bibfnamefont {P.}~\bibnamefont
  {Redfern}}, \ and\ \bibinfo {author} {\bibfnamefont {J.}~\bibnamefont
  {Pople}},\ }\href@noop {} {\bibfield  {journal} {\bibinfo  {journal} {J.
  Chem. Phys.}\ }\textbf {\bibinfo {volume} {106}},\ \bibinfo {pages} {1063}
  (\bibinfo {year} {1997})}\BibitemShut {NoStop}%
\bibitem [{\citenamefont {Zheng}\ \emph {et~al.}(2009)\citenamefont {Zheng},
  \citenamefont {Zhao},\ and\ \citenamefont {Truhlar}}]{Zheng09808}%
  \BibitemOpen
  \bibfield  {author} {\bibinfo {author} {\bibfnamefont {J.~J.}\ \bibnamefont
  {Zheng}}, \bibinfo {author} {\bibfnamefont {Y.}~\bibnamefont {Zhao}}, \ and\
  \bibinfo {author} {\bibfnamefont {D.~G.}\ \bibnamefont {Truhlar}},\
  }\href@noop {} {\bibfield  {journal} {\bibinfo  {journal} {J. Chem. Theory
  Comput.}\ }\textbf {\bibinfo {volume} {5}},\ \bibinfo {pages} {808} (\bibinfo
  {year} {2009})}\BibitemShut {NoStop}%
\bibitem [{\citenamefont {Yang}\ \emph
  {et~al.}(2013{\natexlab{a}})\citenamefont {Yang}, \citenamefont {van
  Aggelen}, \citenamefont {Steinmann}, \citenamefont {Peng},\ and\
  \citenamefont {Yang}}]{Yang13174110}%
  \BibitemOpen
  \bibfield  {author} {\bibinfo {author} {\bibfnamefont {Y.}~\bibnamefont
  {Yang}}, \bibinfo {author} {\bibfnamefont {H.}~\bibnamefont {van Aggelen}},
  \bibinfo {author} {\bibfnamefont {S.~N.}\ \bibnamefont {Steinmann}}, \bibinfo
  {author} {\bibfnamefont {D.}~\bibnamefont {Peng}}, \ and\ \bibinfo {author}
  {\bibfnamefont {W.}~\bibnamefont {Yang}},\ }\href@noop {} {\bibfield
  {journal} {\bibinfo  {journal} {J. Chem. Phys.}\ }\textbf {\bibinfo {volume}
  {139}},\ \bibinfo {pages} {174110} (\bibinfo {year}
  {2013}{\natexlab{a}})}\BibitemShut {NoStop}%
\bibitem [{\citenamefont {Merkle}\ \emph {et~al.}(1992)\citenamefont {Merkle},
  \citenamefont {Savin},\ and\ \citenamefont {Preuss}}]{Merkle929216}%
  \BibitemOpen
  \bibfield  {author} {\bibinfo {author} {\bibfnamefont {R.}~\bibnamefont
  {Merkle}}, \bibinfo {author} {\bibfnamefont {A.}~\bibnamefont {Savin}}, \
  and\ \bibinfo {author} {\bibfnamefont {H.}~\bibnamefont {Preuss}},\
  }\href@noop {} {\bibfield  {journal} {\bibinfo  {journal} {J. Chem. Phys.}\
  }\textbf {\bibinfo {volume} {97}},\ \bibinfo {pages} {9216} (\bibinfo {year}
  {1992})}\BibitemShut {NoStop}%
\bibitem [{\citenamefont {Perdew}\ and\ \citenamefont
  {Zunger}(1981)}]{Perdew815048}%
  \BibitemOpen
  \bibfield  {author} {\bibinfo {author} {\bibfnamefont {J.~P.}\ \bibnamefont
  {Perdew}}\ and\ \bibinfo {author} {\bibfnamefont {A.}~\bibnamefont
  {Zunger}},\ }\href@noop {} {\bibfield  {journal} {\bibinfo  {journal} {Phys.
  Rev. B}\ }\textbf {\bibinfo {volume} {23}},\ \bibinfo {pages} {5048}
  (\bibinfo {year} {1981})}\BibitemShut {NoStop}%
\bibitem [{\citenamefont {Mori-S\'anchez}\ \emph {et~al.}(2008)\citenamefont
  {Mori-S\'anchez}, \citenamefont {Cohen},\ and\ \citenamefont
  {Yang}}]{Mori-Sanchez08146401}%
  \BibitemOpen
  \bibfield  {author} {\bibinfo {author} {\bibfnamefont {P.}~\bibnamefont
  {Mori-S\'anchez}}, \bibinfo {author} {\bibfnamefont {A.~J.}\ \bibnamefont
  {Cohen}}, \ and\ \bibinfo {author} {\bibfnamefont {W.~T.}\ \bibnamefont
  {Yang}},\ }\href@noop {} {\bibfield  {journal} {\bibinfo  {journal} {Phys.
  Rev. Lett.}\ }\textbf {\bibinfo {volume} {100}},\ \bibinfo {pages} {146401}
  (\bibinfo {year} {2008})}\BibitemShut {NoStop}%
\bibitem [{\citenamefont {Cohen}\ \emph {et~al.}(2008)\citenamefont {Cohen},
  \citenamefont {Mori-S\'anchez},\ and\ \citenamefont {Yang}}]{Cohen08121104}%
  \BibitemOpen
  \bibfield  {author} {\bibinfo {author} {\bibfnamefont {A.~J.}\ \bibnamefont
  {Cohen}}, \bibinfo {author} {\bibfnamefont {P.}~\bibnamefont
  {Mori-S\'anchez}}, \ and\ \bibinfo {author} {\bibfnamefont {W.~T.}\
  \bibnamefont {Yang}},\ }\href@noop {} {\bibfield  {journal} {\bibinfo
  {journal} {J. Chem. Phys.}\ }\textbf {\bibinfo {volume} {129}},\ \bibinfo
  {pages} {121104} (\bibinfo {year} {2008})}\BibitemShut {NoStop}%
\bibitem [{\citenamefont {Gori-Giorgi}\ \emph {et~al.}(2009)\citenamefont
  {Gori-Giorgi}, \citenamefont {Seidl},\ and\ \citenamefont
  {Vignale}}]{GoriGiorgi09166402}%
  \BibitemOpen
  \bibfield  {author} {\bibinfo {author} {\bibfnamefont {P.}~\bibnamefont
  {Gori-Giorgi}}, \bibinfo {author} {\bibfnamefont {M.}~\bibnamefont {Seidl}},
  \ and\ \bibinfo {author} {\bibfnamefont {G.}~\bibnamefont {Vignale}},\
  }\href@noop {} {\bibfield  {journal} {\bibinfo  {journal} {Phys. Rev. Lett.}\
  }\textbf {\bibinfo {volume} {{103}}},\ \bibinfo {pages} {{166402}} (\bibinfo
  {year} {{2009}})}\BibitemShut {NoStop}%
\bibitem [{\citenamefont {Gori-Giorgi}\ and\ \citenamefont
  {Seidl}(2010)}]{GoriGiorgi1014405}%
  \BibitemOpen
  \bibfield  {author} {\bibinfo {author} {\bibfnamefont {P.}~\bibnamefont
  {Gori-Giorgi}}\ and\ \bibinfo {author} {\bibfnamefont {M.}~\bibnamefont
  {Seidl}},\ }\href@noop {} {\bibfield  {journal} {\bibinfo  {journal} {Phys.
  Chem. Chem. Phys.}\ }\textbf {\bibinfo {volume} {12}},\ \bibinfo {pages}
  {14405} (\bibinfo {year} {2010})}\BibitemShut {NoStop}%
\bibitem [{\citenamefont {Mirtschink}\ \emph {et~al.}(2013)\citenamefont
  {Mirtschink}, \citenamefont {Seidl},\ and\ \citenamefont
  {Gori-Giorgi}}]{mirtschink_derivative_2013}%
  \BibitemOpen
  \bibfield  {author} {\bibinfo {author} {\bibfnamefont {A.}~\bibnamefont
  {Mirtschink}}, \bibinfo {author} {\bibfnamefont {M.}~\bibnamefont {Seidl}}, \
  and\ \bibinfo {author} {\bibfnamefont {P.}~\bibnamefont {Gori-Giorgi}},\
  }\href@noop {} {\bibfield  {journal} {\bibinfo  {journal} {Phys. Rev. Lett.}\
  }\textbf {\bibinfo {volume} {111}},\ \bibinfo {pages} {126402} (\bibinfo
  {year} {2013})}\BibitemShut {NoStop}%
\bibitem [{\citenamefont {Nafziger}\ and\ \citenamefont
  {Wasserman}(2013)}]{nafziger_delocalization_2013}%
  \BibitemOpen
  \bibfield  {author} {\bibinfo {author} {\bibfnamefont {J.}~\bibnamefont
  {Nafziger}}\ and\ \bibinfo {author} {\bibfnamefont {A.}~\bibnamefont
  {Wasserman}},\ }\href@noop {} {\bibfield  {journal} {\bibinfo  {journal}
  {{arXiv} preprint {arXiv:1305.4966}}\ } (\bibinfo {year} {2013})}\BibitemShut
  {NoStop}%
\bibitem [{\citenamefont {Wagner}\ \emph {et~al.}(2012)\citenamefont {Wagner},
  \citenamefont {Stoudenmire}, \citenamefont {Burke},\ and\ \citenamefont
  {White}}]{WSBW12}%
  \BibitemOpen
  \bibfield  {author} {\bibinfo {author} {\bibfnamefont {L.~O.}\ \bibnamefont
  {Wagner}}, \bibinfo {author} {\bibfnamefont {E.}~\bibnamefont {Stoudenmire}},
  \bibinfo {author} {\bibfnamefont {K.}~\bibnamefont {Burke}}, \ and\ \bibinfo
  {author} {\bibfnamefont {S.~R.}\ \bibnamefont {White}},\ }\href@noop {}
  {\bibfield  {journal} {\bibinfo  {journal} {Phys. Chem. Chem. Phys.}\
  }\textbf {\bibinfo {volume} {14}},\ \bibinfo {pages} {8581 } (\bibinfo {year}
  {2012})}\BibitemShut {NoStop}%
\bibitem [{\citenamefont {van Aggelen}\ \emph {et~al.}(2013)\citenamefont {van
  Aggelen}, \citenamefont {Yang},\ and\ \citenamefont
  {Yang}}]{vanAggelen13030501}%
  \BibitemOpen
  \bibfield  {author} {\bibinfo {author} {\bibfnamefont {H.}~\bibnamefont {van
  Aggelen}}, \bibinfo {author} {\bibfnamefont {Y.}~\bibnamefont {Yang}}, \ and\
  \bibinfo {author} {\bibfnamefont {W.}~\bibnamefont {Yang}},\ }\href@noop {}
  {\bibfield  {journal} {\bibinfo  {journal} {Phys. Rev. A}\ }\textbf {\bibinfo
  {volume} {88}},\ \bibinfo {pages} {030501} (\bibinfo {year}
  {2013})}\BibitemShut {NoStop}%
\bibitem [{\citenamefont {Peng}\ \emph {et~al.}(2013)\citenamefont {Peng},
  \citenamefont {Steinmann}, \citenamefont {van Aggelen},\ and\ \citenamefont
  {Yang}}]{peng_equivalence_2013}%
  \BibitemOpen
  \bibfield  {author} {\bibinfo {author} {\bibfnamefont {D.}~\bibnamefont
  {Peng}}, \bibinfo {author} {\bibfnamefont {S.~N.}\ \bibnamefont {Steinmann}},
  \bibinfo {author} {\bibfnamefont {H.}~\bibnamefont {van Aggelen}}, \ and\
  \bibinfo {author} {\bibfnamefont {W.}~\bibnamefont {Yang}},\ }\href@noop {}
  {\bibfield  {journal} {\bibinfo  {journal} {J. Chem. Phys.}\ }\textbf
  {\bibinfo {volume} {139}},\ \bibinfo {pages} {104112} (\bibinfo {year}
  {2013})}\BibitemShut {NoStop}%
\bibitem [{\citenamefont {Scuseria}\ \emph {et~al.}(2013)\citenamefont
  {Scuseria}, \citenamefont {Henderson},\ and\ \citenamefont
  {Bulik}}]{ScuseriappRPACCSD}%
  \BibitemOpen
  \bibfield  {author} {\bibinfo {author} {\bibfnamefont {G.~E.}\ \bibnamefont
  {Scuseria}}, \bibinfo {author} {\bibfnamefont {T.~M.}\ \bibnamefont
  {Henderson}}, \ and\ \bibinfo {author} {\bibfnamefont {I.~W.}\ \bibnamefont
  {Bulik}},\ }\href@noop {} {\bibfield  {journal} {\bibinfo  {journal} {J.
  Chem. Phys.}\ }\textbf {\bibinfo {volume} {139}},\ \bibinfo {pages} {104113}
  (\bibinfo {year} {2013})}\BibitemShut {NoStop}%
\bibitem [{\citenamefont {Becke}(2005)}]{Becke0564101}%
  \BibitemOpen
  \bibfield  {author} {\bibinfo {author} {\bibfnamefont {A.~D.}\ \bibnamefont
  {Becke}},\ }\href@noop {} {\bibfield  {journal} {\bibinfo  {journal} {J.
  Chem. Phys.}\ }\textbf {\bibinfo {volume} {122}},\ \bibinfo {pages} {064101}
  (\bibinfo {year} {2005})}\BibitemShut {NoStop}%
\bibitem [{\citenamefont {Becke}(2013)}]{becke_density_2013}%
  \BibitemOpen
  \bibfield  {author} {\bibinfo {author} {\bibfnamefont {A.~D.}\ \bibnamefont
  {Becke}},\ }\href@noop {} {\bibfield  {journal} {\bibinfo  {journal} {J.
  Chem. Phys.}\ }\textbf {\bibinfo {volume} {138}},\ \bibinfo {pages} {074109}
  (\bibinfo {year} {2013})}\BibitemShut {NoStop}%
\bibitem [{\citenamefont {Savin}(1988)}]{Savin8859}%
  \BibitemOpen
  \bibfield  {author} {\bibinfo {author} {\bibfnamefont {A.}~\bibnamefont
  {Savin}},\ }\href@noop {} {\bibfield  {journal} {\bibinfo  {journal} {Int. J.
  Quant. Chem. Symp.}\ }\textbf {\bibinfo {volume} {22}},\ \bibinfo {pages}
  {59} (\bibinfo {year} {1988})}\BibitemShut {NoStop}%
\bibitem [{\citenamefont {Grafenstein}\ and\ \citenamefont
  {Cremer}(2000)}]{Grafenstein00569}%
  \BibitemOpen
  \bibfield  {author} {\bibinfo {author} {\bibfnamefont {J.}~\bibnamefont
  {Grafenstein}}\ and\ \bibinfo {author} {\bibfnamefont {D.}~\bibnamefont
  {Cremer}},\ }\href@noop {} {\bibfield  {journal} {\bibinfo  {journal} {Chem.
  Phys. Lett.}\ }\textbf {\bibinfo {volume} {316}},\ \bibinfo {pages} {569}
  (\bibinfo {year} {2000})}\BibitemShut {NoStop}%
\bibitem [{\citenamefont {Knizia}\ and\ \citenamefont {Chan}(2012)}]{FCIcode}%
  \BibitemOpen
  \bibfield  {author} {\bibinfo {author} {\bibfnamefont {G.}~\bibnamefont
  {Knizia}}\ and\ \bibinfo {author} {\bibfnamefont {G.~K.~L.}\ \bibnamefont
  {Chan}},\ }\href@noop {} {\bibfield  {journal} {\bibinfo  {journal} {Phys.
  Rev. Lett.}\ }\textbf {\bibinfo {volume} {109}},\ \bibinfo {pages} {186404}
  (\bibinfo {year} {2012})}\BibitemShut {NoStop}%
\bibitem [{\citenamefont {Knizia}\ and\ \citenamefont {Chan}(2013)}]{FCIcode2}%
  \BibitemOpen
  \bibfield  {author} {\bibinfo {author} {\bibfnamefont {G.}~\bibnamefont
  {Knizia}}\ and\ \bibinfo {author} {\bibfnamefont {G.~K.~L.}\ \bibnamefont
  {Chan}},\ }\href@noop {} {\bibfield  {journal} {\bibinfo  {journal} {J. Chem.
  Theory Comput.}\ }\textbf {\bibinfo {volume} {9}},\ \bibinfo {pages} {1428}
  (\bibinfo {year} {2013})}\BibitemShut {NoStop}%
\bibitem [{\citenamefont {Huang}\ \emph {et~al.}(2011)\citenamefont {Huang},
  \citenamefont {Pavone},\ and\ \citenamefont {Carter}}]{Huang11154110}%
  \BibitemOpen
  \bibfield  {author} {\bibinfo {author} {\bibfnamefont {C.}~\bibnamefont
  {Huang}}, \bibinfo {author} {\bibfnamefont {M.}~\bibnamefont {Pavone}}, \
  and\ \bibinfo {author} {\bibfnamefont {E.~A.}\ \bibnamefont {Carter}},\
  }\href@noop {} {\bibfield  {journal} {\bibinfo  {journal} {J. Chem. Phys.}\
  }\textbf {\bibinfo {volume} {134}},\ \bibinfo {pages} {154110} (\bibinfo
  {year} {2011})}\BibitemShut {NoStop}%
\bibitem [{\citenamefont {Libisch}\ \emph {et~al.}(2012)\citenamefont
  {Libisch}, \citenamefont {Huang}, \citenamefont {Liao}, \citenamefont
  {Pavone},\ and\ \citenamefont {Carter}}]{libisch_origin_2012}%
  \BibitemOpen
  \bibfield  {author} {\bibinfo {author} {\bibfnamefont {F.}~\bibnamefont
  {Libisch}}, \bibinfo {author} {\bibfnamefont {C.}~\bibnamefont {Huang}},
  \bibinfo {author} {\bibfnamefont {P.}~\bibnamefont {Liao}}, \bibinfo {author}
  {\bibfnamefont {M.}~\bibnamefont {Pavone}}, \ and\ \bibinfo {author}
  {\bibfnamefont {E.~A.}\ \bibnamefont {Carter}},\ }\href@noop {} {\bibfield
  {journal} {\bibinfo  {journal} {Phys. Rev. Lett.}\ }\textbf {\bibinfo
  {volume} {109}},\ \bibinfo {pages} {198303} (\bibinfo {year}
  {2012})}\BibitemShut {NoStop}%
\bibitem [{\citenamefont {Perdew}\ \emph {et~al.}(1982)\citenamefont {Perdew},
  \citenamefont {Parr}, \citenamefont {Levy},\ and\ \citenamefont
  {Balduz~Jr}}]{Perdew821691}%
  \BibitemOpen
  \bibfield  {author} {\bibinfo {author} {\bibfnamefont {J.~P.}\ \bibnamefont
  {Perdew}}, \bibinfo {author} {\bibfnamefont {R.~G.}\ \bibnamefont {Parr}},
  \bibinfo {author} {\bibfnamefont {M.}~\bibnamefont {Levy}}, \ and\ \bibinfo
  {author} {\bibfnamefont {J.~L.}\ \bibnamefont {Balduz~Jr}},\ }\href@noop {}
  {\bibfield  {journal} {\bibinfo  {journal} {Phys. Rev. Lett.}\ }\textbf
  {\bibinfo {volume} {49}},\ \bibinfo {pages} {1691} (\bibinfo {year}
  {1982})}\BibitemShut {NoStop}%
\bibitem [{\citenamefont {Perdew}\ and\ \citenamefont
  {Levy}(1983)}]{Perdew831884}%
  \BibitemOpen
  \bibfield  {author} {\bibinfo {author} {\bibfnamefont {J.~P.}\ \bibnamefont
  {Perdew}}\ and\ \bibinfo {author} {\bibfnamefont {M.}~\bibnamefont {Levy}},\
  }\href@noop {} {\bibfield  {journal} {\bibinfo  {journal} {Phys. Rev. Lett.}\
  }\textbf {\bibinfo {volume} {51}},\ \bibinfo {pages} {1884} (\bibinfo {year}
  {1983})}\BibitemShut {NoStop}%
\bibitem [{\citenamefont {Mori-S\'anchez}\ \emph {et~al.}(2009)\citenamefont
  {Mori-S\'anchez}, \citenamefont {Cohen},\ and\ \citenamefont
  {Yang}}]{Mori-Sanchez09}%
  \BibitemOpen
  \bibfield  {author} {\bibinfo {author} {\bibfnamefont {P.}~\bibnamefont
  {Mori-S\'anchez}}, \bibinfo {author} {\bibfnamefont {A.~J.}\ \bibnamefont
  {Cohen}}, \ and\ \bibinfo {author} {\bibfnamefont {W.~T.}\ \bibnamefont
  {Yang}},\ }\href@noop {} {\bibfield  {journal} {\bibinfo  {journal} {Phys.
  Rev. Lett.}\ }\textbf {\bibinfo {volume} {102}},\ \bibinfo {pages} {066403}
  (\bibinfo {year} {2009})}\BibitemShut {NoStop}%
\bibitem [{\citenamefont {Gal}\ and\ \citenamefont
  {Geerlings}(2010)}]{Gal1032512}%
  \BibitemOpen
  \bibfield  {author} {\bibinfo {author} {\bibfnamefont {T.}~\bibnamefont
  {Gal}}\ and\ \bibinfo {author} {\bibfnamefont {P.}~\bibnamefont
  {Geerlings}},\ }\href@noop {} {\bibfield  {journal} {\bibinfo  {journal}
  {Phys. Rev. A}\ }\textbf {\bibinfo {volume} {81}},\ \bibinfo {pages} {032512}
  (\bibinfo {year} {2010})}\BibitemShut {NoStop}%
\bibitem [{\citenamefont {Yang}\ \emph
  {et~al.}(2013{\natexlab{b}})\citenamefont {Yang}, \citenamefont
  {Mori-Sánchez},\ and\ \citenamefont {Cohen}}]{yang_extension_2013}%
  \BibitemOpen
  \bibfield  {author} {\bibinfo {author} {\bibfnamefont {W.}~\bibnamefont
  {Yang}}, \bibinfo {author} {\bibfnamefont {P.}~\bibnamefont {Mori-Sánchez}},
  \ and\ \bibinfo {author} {\bibfnamefont {A.~J.}\ \bibnamefont {Cohen}},\
  }\href@noop {} {\bibfield  {journal} {\bibinfo  {journal} {J. Chem. Phys.}\
  }\textbf {\bibinfo {volume} {139}},\ \bibinfo {pages} {104114} (\bibinfo
  {year} {2013}{\natexlab{b}})}\BibitemShut {NoStop}%
\bibitem [{\citenamefont {Steinmann}\ and\ \citenamefont
  {Yang}(2013)}]{Steinmann13074107}%
  \BibitemOpen
  \bibfield  {author} {\bibinfo {author} {\bibfnamefont {S.~N.}\ \bibnamefont
  {Steinmann}}\ and\ \bibinfo {author} {\bibfnamefont {W.}~\bibnamefont
  {Yang}},\ }\href@noop {} {\bibfield  {journal} {\bibinfo  {journal} {J. Chem.
  Phys.}\ }\textbf {\bibinfo {volume} {139}},\ \bibinfo {pages} {074107}
  (\bibinfo {year} {2013})}\BibitemShut {NoStop}%
\bibitem [{\citenamefont {Hubbard}(1963)}]{Hubbard26111963}%
  \BibitemOpen
  \bibfield  {author} {\bibinfo {author} {\bibfnamefont {J.}~\bibnamefont
  {Hubbard}},\ }\href@noop {} {\bibfield  {journal} {\bibinfo  {journal} {Proc.
  R. Soc. A}\ }\textbf {\bibinfo {volume} {276}},\ \bibinfo {pages} {238}
  (\bibinfo {year} {1963})}\BibitemShut {NoStop}%
\bibitem [{\citenamefont {Lieb}\ and\ \citenamefont {Wu}(1968)}]{Lieb681445}%
  \BibitemOpen
  \bibfield  {author} {\bibinfo {author} {\bibfnamefont {E.~H.}\ \bibnamefont
  {Lieb}}\ and\ \bibinfo {author} {\bibfnamefont {F.~Y.}\ \bibnamefont {Wu}},\
  }\href@noop {} {\bibfield  {journal} {\bibinfo  {journal} {Phys. Rev. Lett.}\
  }\textbf {\bibinfo {volume} {20}},\ \bibinfo {pages} {1445} (\bibinfo {year}
  {1968})}\BibitemShut {NoStop}%
\bibitem [{\citenamefont {Lima}\ \emph {et~al.}(2003)\citenamefont {Lima},
  \citenamefont {Silva}, \citenamefont {Oliveira},\ and\ \citenamefont
  {Capelle}}]{CapelleHubbard}%
  \BibitemOpen
  \bibfield  {author} {\bibinfo {author} {\bibfnamefont {N.~A.}\ \bibnamefont
  {Lima}}, \bibinfo {author} {\bibfnamefont {M.~F.}\ \bibnamefont {Silva}},
  \bibinfo {author} {\bibfnamefont {L.~N.}\ \bibnamefont {Oliveira}}, \ and\
  \bibinfo {author} {\bibfnamefont {K.}~\bibnamefont {Capelle}},\ }\href@noop
  {} {\bibfield  {journal} {\bibinfo  {journal} {Phys. Rev. Lett.}\ }\textbf
  {\bibinfo {volume} {90}},\ \bibinfo {pages} {146402} (\bibinfo {year}
  {2003})}\BibitemShut {NoStop}%
\bibitem [{\citenamefont {França}\ \emph {et~al.}(2012)\citenamefont {França},
  \citenamefont {Vieira},\ and\ \citenamefont {Capelle}}]{Franca12073021}%
  \BibitemOpen
  \bibfield  {author} {\bibinfo {author} {\bibfnamefont {V.~V.}\ \bibnamefont
  {França}}, \bibinfo {author} {\bibfnamefont {D.}~\bibnamefont {Vieira}}, \
  and\ \bibinfo {author} {\bibfnamefont {K.}~\bibnamefont {Capelle}},\
  }\href@noop {} {\bibfield  {journal} {\bibinfo  {journal} {New J. Phys.}\
  }\textbf {\bibinfo {volume} {14}},\ \bibinfo {pages} {073021} (\bibinfo
  {year} {2012})}\BibitemShut {NoStop}%
\bibitem [{\citenamefont {Capelle}\ and\ \citenamefont
  {Campo}(2013)}]{CapelleReview}%
  \BibitemOpen
  \bibfield  {author} {\bibinfo {author} {\bibfnamefont {K.}~\bibnamefont
  {Capelle}}\ and\ \bibinfo {author} {\bibfnamefont {V.~L.}\ \bibnamefont
  {Campo}, \bibfnamefont {Jr}},\ }\href@noop {} {\bibfield  {journal} {\bibinfo
   {journal} {Phys. Rep.}\ }\textbf {\bibinfo {volume} {528}},\ \bibinfo
  {pages} {91} (\bibinfo {year} {2013})}\BibitemShut {NoStop}%
\bibitem [{\citenamefont {Mori-S\'anchez}\ \emph {et~al.}(2012)\citenamefont
  {Mori-S\'anchez}, \citenamefont {Cohen},\ and\ \citenamefont
  {Yang}}]{Mori-Sanchez12042507}%
  \BibitemOpen
  \bibfield  {author} {\bibinfo {author} {\bibfnamefont {P.}~\bibnamefont
  {Mori-S\'anchez}}, \bibinfo {author} {\bibfnamefont {A.~J.}\ \bibnamefont
  {Cohen}}, \ and\ \bibinfo {author} {\bibfnamefont {W.~T.}\ \bibnamefont
  {Yang}},\ }\href@noop {} {\bibfield  {journal} {\bibinfo  {journal} {Phys.
  Rev. A}\ }\textbf {\bibinfo {volume} {85}},\ \bibinfo {pages} {042507}
  (\bibinfo {year} {2012})}\BibitemShut {NoStop}%
\bibitem [{\citenamefont {Cohen}\ and\ \citenamefont
  {Mori-Sánchez}(2014)}]{cohen_dramatic_2014}%
  \BibitemOpen
  \bibfield  {author} {\bibinfo {author} {\bibfnamefont {A.~J.}\ \bibnamefont
  {Cohen}}\ and\ \bibinfo {author} {\bibfnamefont {P.}~\bibnamefont
  {Mori-Sánchez}},\ }\href@noop {} {\bibfield  {journal} {\bibinfo  {journal}
  {J. Chem. Phys.}\ }\textbf {\bibinfo {volume} {140}},\ \bibinfo {pages}
  {044110} (\bibinfo {year} {2014})}\BibitemShut {NoStop}%
\bibitem [{\citenamefont {Mori-S\'anchez}\ \emph {et~al.}(2014)\citenamefont
  {Mori-S\'anchez}, \citenamefont {Yang},\ and\ \citenamefont {Cohen}}]{GGKS}%
  \BibitemOpen
  \bibfield  {author} {\bibinfo {author} {\bibfnamefont {P.}~\bibnamefont
  {Mori-S\'anchez}}, \bibinfo {author} {\bibfnamefont {W.~T.}\ \bibnamefont
  {Yang}}, \ and\ \bibinfo {author} {\bibfnamefont {A.~J.}\ \bibnamefont
  {Cohen}},\ }\href@noop {} {\  (\bibinfo {year} {2014})},\ \bibinfo {note}
  {manuscript in preparation}\BibitemShut {NoStop}%
\bibitem [{\citenamefont {Perdew}(1985)}]{perdew_what_1985}%
  \BibitemOpen
  \bibfield  {author} {\bibinfo {author} {\bibfnamefont {J.~P.}\ \bibnamefont
  {Perdew}},\ }in\ \href@noop {} {\emph {\bibinfo {booktitle} {Density
  Functional Methods in Physics}}},\ \bibinfo {editor} {edited by\ \bibinfo
  {editor} {\bibfnamefont {R.~M.}\ \bibnamefont {Dreizler}}\ and\ \bibinfo
  {editor} {\bibnamefont {Providencia}}}\ (\bibinfo {year} {1985})\ pp.\
  \bibinfo {pages} {265--308}\BibitemShut {NoStop}%
\bibitem [{\citenamefont {Gritsenko}\ \emph {et~al.}(1995)\citenamefont
  {Gritsenko}, \citenamefont {Vanleeuwen},\ and\ \citenamefont
  {Baerends}}]{Gritsenko951870}%
  \BibitemOpen
  \bibfield  {author} {\bibinfo {author} {\bibfnamefont {O.~V.}\ \bibnamefont
  {Gritsenko}}, \bibinfo {author} {\bibfnamefont {R.}~\bibnamefont
  {Vanleeuwen}}, \ and\ \bibinfo {author} {\bibfnamefont {E.~J.}\ \bibnamefont
  {Baerends}},\ }\href@noop {} {\bibfield  {journal} {\bibinfo  {journal}
  {Phys. Rev. A}\ }\textbf {\bibinfo {volume} {52}},\ \bibinfo {pages} {1870}
  (\bibinfo {year} {1995})}\BibitemShut {NoStop}%
\bibitem [{\citenamefont {Helbig}\ \emph {et~al.}(2009)\citenamefont {Helbig},
  \citenamefont {Tokatly},\ and\ \citenamefont {Rubio}}]{Helbig09224105}%
  \BibitemOpen
  \bibfield  {author} {\bibinfo {author} {\bibfnamefont {N.}~\bibnamefont
  {Helbig}}, \bibinfo {author} {\bibfnamefont {I.~V.}\ \bibnamefont {Tokatly}},
  \ and\ \bibinfo {author} {\bibfnamefont {A.}~\bibnamefont {Rubio}},\
  }\href@noop {} {\bibfield  {journal} {\bibinfo  {journal} {J. Chem. Phys.}\
  }\textbf {\bibinfo {volume} {131}},\ \bibinfo {pages} {224105} (\bibinfo
  {year} {2009})}\BibitemShut {NoStop}%
\bibitem [{\citenamefont {Tempel}\ \emph {et~al.}(2009)\citenamefont {Tempel},
  \citenamefont {Martinez},\ and\ \citenamefont
  {Maitra}}]{tempel_revisiting_2009}%
  \BibitemOpen
  \bibfield  {author} {\bibinfo {author} {\bibfnamefont {D.~G.}\ \bibnamefont
  {Tempel}}, \bibinfo {author} {\bibfnamefont {T.~J.}\ \bibnamefont
  {Martinez}}, \ and\ \bibinfo {author} {\bibfnamefont {N.~T.}\ \bibnamefont
  {Maitra}},\ }\href@noop {} {\bibfield  {journal} {\bibinfo  {journal} {J.
  Chem. Theory Comput.}\ }\textbf {\bibinfo {volume} {5}},\ \bibinfo {pages}
  {770 } (\bibinfo {year} {2009})}\BibitemShut {NoStop}%
\bibitem [{\citenamefont {Hellgren}\ and\ \citenamefont
  {Gross}(2012)}]{Hellgren12022514}%
  \BibitemOpen
  \bibfield  {author} {\bibinfo {author} {\bibfnamefont {M.}~\bibnamefont
  {Hellgren}}\ and\ \bibinfo {author} {\bibfnamefont {E.~K.~U.}\ \bibnamefont
  {Gross}},\ }\href@noop {} {\bibfield  {journal} {\bibinfo  {journal} {Phys.
  Rev. A}\ }\textbf {\bibinfo {volume} {85}},\ \bibinfo {pages} {022514}
  (\bibinfo {year} {2012})}\BibitemShut {NoStop}%
\bibitem [{\citenamefont {Johnson}\ \emph {et~al.}(2010)\citenamefont
  {Johnson}, \citenamefont {Keinan}, \citenamefont {Mori-S\'anchez},
  \citenamefont {Contreras-Garcia}, \citenamefont {Cohen},\ and\ \citenamefont
  {Yang}}]{Johnson106498}%
  \BibitemOpen
  \bibfield  {author} {\bibinfo {author} {\bibfnamefont {E.~R.}\ \bibnamefont
  {Johnson}}, \bibinfo {author} {\bibfnamefont {S.}~\bibnamefont {Keinan}},
  \bibinfo {author} {\bibfnamefont {P.}~\bibnamefont {Mori-S\'anchez}},
  \bibinfo {author} {\bibfnamefont {J.}~\bibnamefont {Contreras-Garcia}},
  \bibinfo {author} {\bibfnamefont {A.~J.}\ \bibnamefont {Cohen}}, \ and\
  \bibinfo {author} {\bibfnamefont {W.~T.}\ \bibnamefont {Yang}},\ }\href@noop
  {} {\bibfield  {journal} {\bibinfo  {journal} {J. Am. Chem. Soc.}\ }\textbf
  {\bibinfo {volume} {132}},\ \bibinfo {pages} {6498} (\bibinfo {year}
  {2010})}\BibitemShut {NoStop}%
\bibitem [{\citenamefont {Klimo}\ and\ \citenamefont
  {Ti\v{n}o}(1981)}]{klimo_study_1981}%
  \BibitemOpen
  \bibfield  {author} {\bibinfo {author} {\bibfnamefont {V.}~\bibnamefont
  {Klimo}}\ and\ \bibinfo {author} {\bibfnamefont {J.}~\bibnamefont
  {Ti\v{n}o}},\ }\href@noop {} {\bibfield  {journal} {\bibinfo  {journal}
  {Collect. Czech. Chem. Commun.}\ }\textbf {\bibinfo {volume} {46}},\ \bibinfo
  {pages} {1365 } (\bibinfo {year} {1981})}\BibitemShut {NoStop}%
\bibitem [{\citenamefont {Nobes}\ \emph {et~al.}(1991)\citenamefont {Nobes},
  \citenamefont {Moncrieff}, \citenamefont {Wong}, \citenamefont {Radom},
  \citenamefont {Gill},\ and\ \citenamefont {Pople}}]{Nobes1991216}%
  \BibitemOpen
  \bibfield  {author} {\bibinfo {author} {\bibfnamefont {R.~H.}\ \bibnamefont
  {Nobes}}, \bibinfo {author} {\bibfnamefont {D.}~\bibnamefont {Moncrieff}},
  \bibinfo {author} {\bibfnamefont {M.~W.}\ \bibnamefont {Wong}}, \bibinfo
  {author} {\bibfnamefont {L.}~\bibnamefont {Radom}}, \bibinfo {author}
  {\bibfnamefont {P.~M.~W.}\ \bibnamefont {Gill}}, \ and\ \bibinfo {author}
  {\bibfnamefont {J.~A.}\ \bibnamefont {Pople}},\ }\href@noop {} {\bibfield
  {journal} {\bibinfo  {journal} {Chem. Phys. Lett.}\ }\textbf {\bibinfo
  {volume} {182}},\ \bibinfo {pages} {216 } (\bibinfo {year}
  {1991})}\BibitemShut {NoStop}%
\bibitem [{\citenamefont {Perdew}\ \emph {et~al.}(1995)\citenamefont {Perdew},
  \citenamefont {Savin},\ and\ \citenamefont {Burke}}]{Perdew954531}%
  \BibitemOpen
  \bibfield  {author} {\bibinfo {author} {\bibfnamefont {J.~P.}\ \bibnamefont
  {Perdew}}, \bibinfo {author} {\bibfnamefont {A.}~\bibnamefont {Savin}}, \
  and\ \bibinfo {author} {\bibfnamefont {K.}~\bibnamefont {Burke}},\
  }\href@noop {} {\bibfield  {journal} {\bibinfo  {journal} {Phys. Rev. A}\
  }\textbf {\bibinfo {volume} {51}},\ \bibinfo {pages} {4531} (\bibinfo {year}
  {1995})}\BibitemShut {NoStop}%
\bibitem [{\citenamefont {Stein}\ \emph {et~al.}(2014)\citenamefont {Stein},
  \citenamefont {Jiménez-Hoyos},\ and\ \citenamefont
  {Scuseria}}]{stein_stability_2014}%
  \BibitemOpen
  \bibfield  {author} {\bibinfo {author} {\bibfnamefont {T.}~\bibnamefont
  {Stein}}, \bibinfo {author} {\bibfnamefont {C.~A.}\ \bibnamefont
  {Jiménez-Hoyos}}, \ and\ \bibinfo {author} {\bibfnamefont {G.~E.}\
  \bibnamefont {Scuseria}},\ }\href@noop {} {\bibfield  {journal} {\bibinfo
  {journal} {J. Phys. Chem. A}\ ,\ \bibinfo {pages} {In~Press}} (\bibinfo
  {year} {2014})}\BibitemShut {NoStop}%
\end{thebibliography}
%

\end{document}